\documentclass{aa} 
\usepackage{graphicx}
\usepackage{caption}
\usepackage{subcaption}
\usepackage{txfonts}
\usepackage{xcolor}
\usepackage[normalem]{ulem}
\usepackage{amsmath}
\usepackage{multirow}
\usepackage{booktabs}
\usepackage{bm}
\usepackage{arydshln}
\usepackage[breaklinks=true]{hyperref}

\begin{document} 

\defcitealias{paperI}{Paper I}

   \title{A hierarchical Bayesian model to infer \textit{PL(Z)} relations using
     \textit{Gaia} parallaxes}

   \author{
     H.E. Delgado \inst{1}
     \and 
     L.M. Sarro \inst{1}
     \and
     G. Clementini \inst{2}
     \and
     T. Muraveva \inst{2}
     \and
     A. Garofalo \inst{2,3}
   }

   \institute{Dpto. de Inteligencia Artificial, UNED, c/
	     Juan del Rosal, 16, 28040 Madrid, Spain. 
    \and INAF,
	     Osservatorio di Astrofisica e Scienza dello Spazio di Bologna, via Piero Gobetti 93/3,
	     40129 Bologna, Italy.
	\and Dipartimento di Fisica e Astronomia, Università di Bologna, via
	     Piero Gobetti 93/2, 40129 Bologna, Italy.}

   \date{Received \today; accepted}

  \abstract
  { \cite{paperIII} analysed Period-Luminosity-Metallicity ($PLZ$) 
  	relations for RR~Lyrae stars using the \textit{Gaia} Data Release 
  	2 parallaxes. It built on a previous work presented in \cite{paperI} 
  	that was based on the first {\textit Gaia} Data Release, and also 
  	included Period-Luminosity ($PL$) relations for Cepheids and 
  	RR~Lyrae stars. The method used to infer the relations in
    \cite{paperIII} \citep[and one of the methods used in][]{paperI}
    was based on a hierarchical Bayesian model, the full description
    of which was deferred to a subsequent publication that is
    presented here.}
  {We aim at creating a Bayesian model to infer the coefficients of
    $PL$ or $PLZ$ relations that propagates uncertainties in the
    observables in a rigorous and well founded way.}
  {We propose a directed acyclic graph to encode the conditional
    probabilities of the inference model that will allow us to infer
    probability distributions for the $PL$ and $PL(Z)$ relations. We
    evaluate the model with several semi-synthetic data sets and apply
    it to a sample of 200 fundamental mode and first overtone mode RR
    Lyrae stars for which Gaia DR1 parallaxes and literature $K_{\rm s}$-band
    mean magnitudes are available. We define and test several
    hyperprior probabilities to verify their adequacy and check the
    sensitivity of the solution with respect to the prior choice.}
  {The main conclusion of this work is the absolute necessity of
    incorporating the existing correlations between the observed
    variables (periods, metallicities and parallaxes) in the form of
    model priors in order to avoid systematically biased results,
    especially in the case of non-negligible uncertainties in the
    parallaxes. The tests with the semi-synthetic data based on the
    data set used in \cite{paperI} reveal the significant impact that
    the existing correlations between parallax, metallicity and
    periods have on the inferred parameters. The relation coefficients
    obtained here have been superseded by those presented in
    \cite{paperIII}, that incorporates the findings of this work and
    the more recent \textit{Gaia} DR2 measurements.}
   {}


   \maketitle
   \section{Introduction}
   \label{sec:intro}

Cepheids and RR Lyrae stars are primary standard candles of the
cosmological distance ladder because they follow canonical relations
that for Cepheids link the star intrinsic luminosity ($L$) to the
period ($P$) of light variation, traditionally referred to as 
period-luminosity relation or Leavitt Law { (\citealt{Leavitt-1912}; 
\citealt{Madore-Fredman-1991}; \citealt{Freedman-et-al-2001}; 
\citealt{Marconi-et-al-2005}; \citealt{Saha-et-al-2006}; 
\citealt{Riess-et-al-2011,Riess-et-al-2016,Riess-at-al-2018}; 
\citealt{Ripepi-et-al-2012}; \citealt{Gieren-et-al-2013}; \citealt{paperI}, 
hereafter Paper I,  and references therein) whereas for RR
Lyrae stars link $L$ in the infrared passbands to $P$ and possibly the
stellar metallicity ($Z$; $PL$ - metallicity relation -- $PL(Z)$;  
\citealt{Longmore-et-al-1986}; \citealt{Solima-et-al-2006,Solima-et-al-2008};
\citealt{Borissova-et-al-2009}; \citealt{Marconi-et-al-2015}; 
\citealt{Neeley-et-al-2017}; \citealt{Sesar-et-al-2017}; 
\citealt{paperIII,Muraveva-et-al-2018b}; \citetalias{paperI} and references therein)
or $L$ in the visual passband to $Z$ (traditionally referred to as RR
Lyrae luminosity - metallicity relation; \citealt{Cacciari-Clementini-2003}, 
\citealt{Clementini-et-al-2003}, \citealt{Bono-2003}, \citealt{Catelan-2004}, 
\citetalias{paperI}, \citealt{paperIII} and references therein). }  The predicted 
precision of the \textit{Gaia} end-of-mission parallaxes for local 
Cepheids and RR Lyrae stars\footnote{See
  \url{https://www.cosmos.esa.int/web/gaia/science-performance}} will
allow us to determine the slope and zero point of these fundamental
relations with unprecedented accuracy, thus setting the basis for a
global reassessment of the whole cosmic distance ladder.  As a first
anticipation of the \textit{Gaia} potential in this field of the
cosmic distance ladder and a first assessment of improved precision
with respect to previous astrometric missions such as, for instance,
Hipparcos, and the dramatic increase in statistics compared to what is
achievable for instance, measuring parallaxes with the Hubble Space
Telescope, \textit{Gaia} DR1 published parallaxes for more than 700
Galactic Cepheids and RR Lyrae stars, computed as part of the
Tycho-\textit{Gaia} Astrometric Solution (TGAS;
\cite{Lindegren-et-al-2016}). { A number of papers after \textit{Gaia} 
intermediate data releases in 2016 and 2018 have addressed Cepheids 
and RR Lyrae stars (e.g. \citealt{Lindegren-et-al-2016}; 
\citealt{Clementini-et-al-2016, Clementini-et-al-2018}; 
\citealt{Arenou-et-al-2017,Arenou-et-al-2018}) with specific emphasis 
in their use as standard candles (\citealt{Casertano-et-al-2017}; 
\citetalias{paperI}; \citealt{Riess-at-al-2018}; \citealt{paperIII}).}
In \citetalias{paperI} we have used TGAS parallaxes, along with 
literature photometry and spectroscopy, to calibrate the zero 
point of the $PL$ relations of classical and type II Cepheids, 
and the near-infrared $PL$ and $PL(Z)$
relations of RR Lyrae stars by fitting these relations adopting
different techniques that operate either in parallax or absolute
magnitude space. In that paper different sources of biases affecting
the TGAS samples of Cepheids and RR Lyrae stars were discussed at some
length and the possible systematic errors caused in the inferred
luminosity calibrations were analysed in detail.
   
Section 3.2 of \citetalias{paperI} in particular discussed the problem
of fitting general luminosity relations between the absolute magnitude
$M_{\rm True}$, the decadic logarithm of the period $P_{\rm True} $
and possibly also the metallicity $\left[\rm{Fe/H}\right]_{\rm{True}}$
of the form

\begin{equation}
  M_{\rm True} = b + c\cdot \log(P_{\rm True}) + k\cdot
  \left[\rm{Fe/H}\right]_{\rm{True}}
  \label{eq:plz-intro}
\end{equation}
   
\noindent with a sample that is truncated in parallax (by removing the
non-positive values) and for which the assumption of normality of
uncertainties in the absolute magnitude is not valid. { A more
  detailed description of the intricacies involved in using
  astrometric measurements for the inference of quantities of
  astrophysical interest in general, and PLZ relation coefficients in
  particular can be found in \cite{Luri-at-al-2018}}. Our proposal
in \citetalias{paperI} was to construct a two-level statistical model
that distinguishes between true and measured parallaxes. { This
  model can then be used to infer the true parallaxes and absolute
  magnitudes from the measurements.}  { One of the rigorous ways} 
to construct such a model is to apply the Bayesian methodology where one assigns a
prior probability distribution to the true parallax population.  In
doing so, a suitable selection of this prior will improve the
estimation of individual true parallaxes in the sense that their
posterior credible intervals are ``shrunken'' with respect to the
measurement uncertainties. Setting a specific prior is always
controversial, but in principle it is possible to define only a
functional form that depends on a set of unknown parameters. The
specific prior is then inferred from the data as part of the global
inference process. This prior functional form should be flexible
enough to properly model the true distribution of parallaxes but also
should be sufficiently restrictive to enforce a plausible distribution
for the true parallaxes on the basis of the knowledge present in the
astronomical literature.
        
{ The solution described in the previous paragraph can be
  represented as a graph model that incorporates the $PL(Z)$ relation,
  the definition of absolute magnitudes in terms of the apparent
  magnitude $m$ and the parallax $\varpi$, and the corresponding
  distribution of the measurements given the true values.} This way we
guarantee that the observational uncertainties are simultaneously
propagated through the graph and that the uncertainties of the
parameters of the $PLZ$ relationship are estimated in a way that is
consistent with the measurement uncertainties. Also, the effect of
including the relationship
    
\begin{equation}
  b+c\cdot\log P+k\cdot \left[\rm{Fe/H}\right]=m + 5\log\varpi-10\, 
  \label{eq:PLZ-pi}
\end{equation}
    
\noindent in the model is to constrain the parameter space in such a
way that the $PLZ$ relationship coefficients and the individual true
parallaxes have to be consistent.
   
The objective of this paper is to infer estimates of the parameters of
the $PLZ$ relationship. We apply the hierarchical Bayesian
methodology, which consists { of} dividing the variability of the
statistical inference problem into several levels. In this way we
partition the parameter space associated { with} inferring the $PLZ$
relation into population-level parameters and observations.  We
represent the hierarchical Bayesian model with a directed acyclic
graph and perform the inference using Markov chain Monte Carlo (MCMC)
simulation techniques \citep{robertCasella2013}.  A minimal
description of the methodology and preliminary results was already
presented in \citetalias{paperI} which we intend to extend and clarify
here.  For reasons of clarity and scope we focus on the inference of
the $PLZ$ relationship in the $K$-band for 200 fundamental and first
overtone RR Lyrae stars, the main properties of which are provided in
Table A.3 of \citetalias{paperI}. The model is applicable with minimal
modifications to other variability types such as Cepheids or Long
Period Variables and different photometric bands. In this work we
present the results of the full model including the slopes of the
relation, expanding the results presented in \citetalias{paperI} where
only the zero points were inferred while the slopes were fixed to
literature values.
   
A similar methodology has been applied by \cite{Sesar-et-al-2017}
to constrain $PLZ$ relations of { fundamental mode (ab type)} 
RR Lyrae stars in the mid-infrared
W1 and W2 bands { of} the Wide-field Infrared Survey Explore (WISE;
\cite{Wright-et-al-2010}), using TGAS parallaxes, but modelling true
distances with an exponentially decreasing volume density (hereafter
EDVD) prior proposed by \cite{Bailer-Jones-2015}.

{ The Bayesian hierarchical method presented in \citetalias{paperI}} used 
a log-normal prior to model the distribution of true parallaxes { 
independently of the other model parameters}. With this prior, the $\log\left(P\right)$
slope turned out to be severely underestimated when compared to the
literature values, although this result was not specifically discussed
therein. In the present work we extend the Bayesian analysis performed
in \citetalias{paperI} in three directions.  { First, we validate
  the model with semi-synthetic data and analyse the causes of the
  slope underestimation. Second, we extend the Bayesian analysis by
  testing alternative prior distributions for parallaxes and
  demonstrate that one of them mitigates to some degree the problem of
  the underestimation of the $PLZ$ $\log\left(P\right)$ slope. Third,
  we study the sensitivity of the Bayesian analysis results under
  different prior choices for some critical hyperparameters of our
  hierarchical model (HM).}
      
The structure of the paper is as follows. In Sect. \ref{sec:hiermodel}
we summarize the theoretical foundations of the hierarchical Bayesian
methodology and describe extensively the HM used for inferring the
$PLZ$ relationship in \citetalias{paperI} { and \cite{paperIII} (in
  the latter case with minor adaptations)}. In
Sect. \ref{sec:validation} { we study the data set in detail and
  explore its properties by means of semi-synthetic samples
  constructed assuming a known PLZ relation;} in Sect. \ref{sec:post}
  we present the full results of the MCMC samples of the posterior
  distribution for the Gaia DR1 data used in \citetalias{paperI};{ in
  Sect. \ref{sec:sensi} we study the sensitivity of the results to the
  choice of hyper-parameters, and in Sect. \ref{sec:summary} we
  summarise the findings of the paper.}

\section{The hierarchical Bayesian model}
\label{sec:hiermodel}

A full introduction to Bayesian inference and hierarchical Bayes is
beyond the scope of this manuscript. We recommend the interested
reader to consult \cite{Gelman2004} and \cite{GelmanHill-2007} for
very pedagogic introductions, and \cite{Luri-at-al-2018} for a
more Astronomy-oriented introduction. In what follows, we summarize
the main concepts of the methodology. Bayesian inference is based on
Bayes' rule:

\begin{equation}
  p\left(\Theta\mid\mathcal{D}\right)\propto
  p\left(\mathcal{D}\mid\Theta\right)\times p\left(\Theta\right)\,,
  \label{eq:bayes2-1}
\end{equation}

\noindent where $\mathcal{D}$ are the observations (data), $\Theta$
are the parameters of a model proposed to explain the data and $p$
represents a probability distribution.  The right side of
Eq. \ref{eq:bayes2-1} represents the model itself, specified by the
joint probability distribution $p\left(\mathcal{D},\Theta\right)$ of
the data and the parameters. This distribution factorizes into:

\begin{itemize}
\item the conditional distribution
  $p\left(\mathcal{D}\mid\Theta\right)$ of the data given the
  parameters (the so called \textit{likelihood}), and
\item the \textit{prior} distribution of the parameters
  $p\left(\Theta\right)$, which represents our knowledge about
  plausible parameter values before observing the data.
\end{itemize}

The basic model of Eq. \ref{eq:bayes2-1} divides the variability of
the statistical problem into two levels: observations and
parameters. The hierarchical Bayesian methodology consists { of}
distinguishing further levels of variability. In our case, we
introduce a new dependence of the prior distribution
$p\left(\Theta\right)$ on a new set of parameters $\Phi$ (the so
called \textit {hyperparameters}) and assign \textit {hyperprior}
distributions $p\left(\Phi\right)$ to them. We explain its nature in
the following sections. In order to have a better understanding of the
dependency structure dictated by the model, it is customary to
represent the factorization of the joint probability distribution
$p\left(D,\mathbf{\Theta},\Phi\right)$ by using the Bayesian network
formalism \citep{Pearl1988,Lauritzen1996}, which consists { of} drawing a
DAG (directed acyclic graph) in which nodes encode model parameters,
measurements or constants, and directed links represent conditional
probability dependence relationships.
   
The inference in a hierarchical Bayesian model proceeds by calculating
the \textit {marginal joint posterior} distribution of a set of
\textit{parameters of interest} given the data. { In complex problems with 
many parameters the posterior distribution usually is not available
in an analytically tractable closed form but can be approximately 
evaluated using MCMC simulation techniques. }

\subsection{{ Conditional dependencies}}
\label{sec:cond_deps}

\begin{figure*}
  \begin{center}
    \hspace*{-0.5cm}
    \includegraphics[scale=.4]{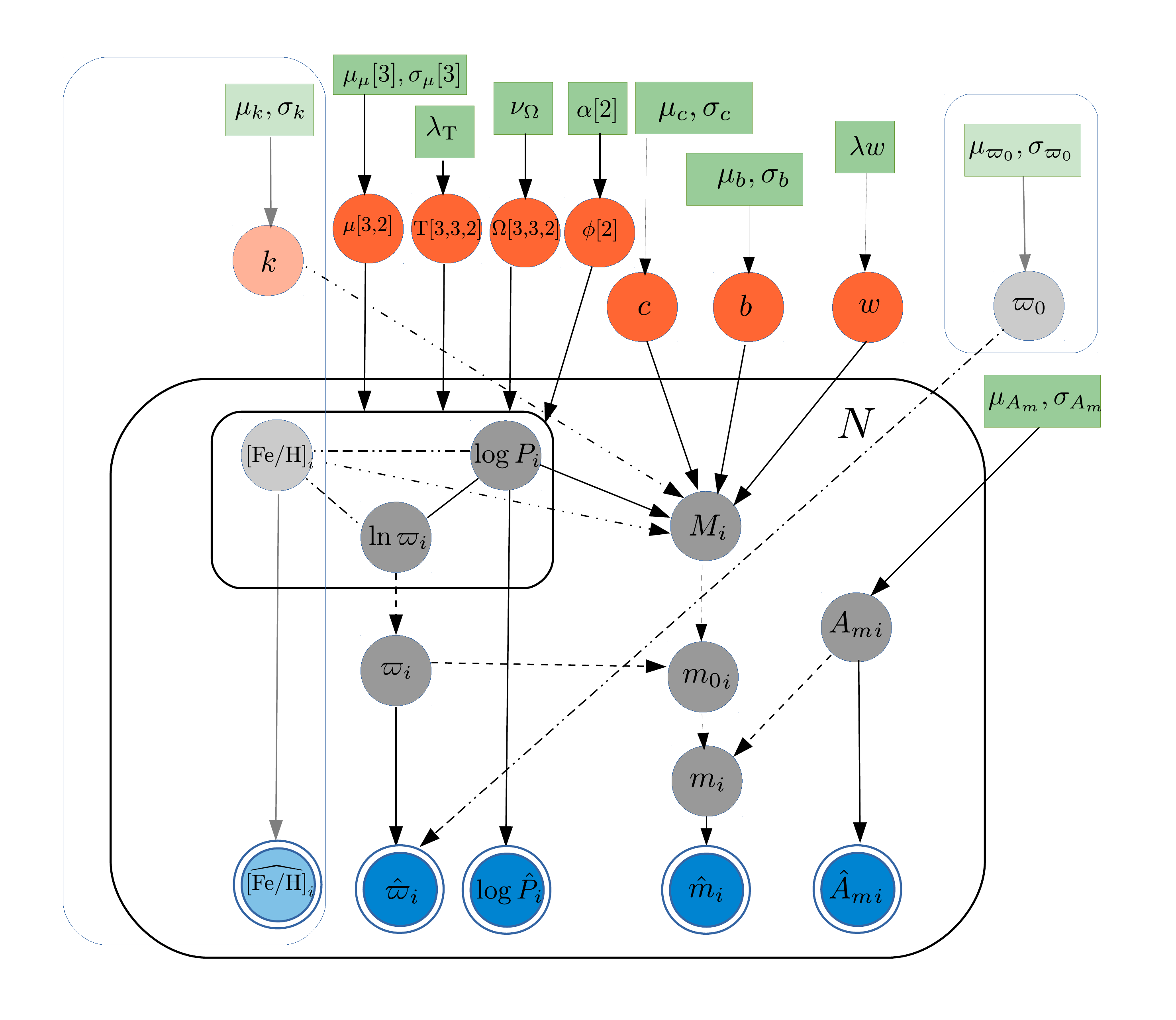}
    \caption{Directed Acyclic Graph that represents the forward model
      used to infer the $PLZ$ relation coefficients { when the prior
      of true metallicities, logarithm of true periods and (natural) logarithm of 
      true parallaxes is assumed to be a 3D Gaussian mixture distribution.}}
    \label{DAG}
  \end{center}
\end{figure*}

In this and the next { section} we describe { our sample} and the
hierarchical model that encodes the conditional probability relations
between the observations and the parameters of the linear $PL(Z)$
relations. We include Fig. 19 of \citetalias{paperI} here as
Fig. \ref{DAG}, to facilitate reading, but include some additional
clarifications that could not be described there due to space and
scope limitations.
      
In the following, we will change the notation to avoid cluttering of
subscripts. Henceforth, we will denote measured quantities with a
circumflex accent (\verb!^!)  and remove the subscript ${\rm True}$
from the true values. The DAG in Fig. \ref{DAG} encodes the
probabilistic relationships between the variables of our model and
shows the measurements at the bottom level: decadic logarithm of
periods $\log\hat{P}_{i}$, apparent magnitudes $\hat{m}_i$,
metallicities $\widehat{\left[\mathrm{Fe/H}\right]}_{i}$, parallaxes
$\hat{\varpi}_i$ and extinctions $\hat{A}_{m_{i}}$ The subindex $i$
runs from 1 to the total number of stars $N$ in each sample. Our model
assumes that the measurements

   \begin{equation}
   \mathbf{d}_{i}=\left(\hat{m}_{i},\log\hat{P}_{i},\hat{\varpi}_{i},
   \widehat{\left[\mathrm{Fe/H}\right]}_{i},
   \hat{A}_{m_{i}}\right)\,, 
   \label{eq:data}
   \end{equation}

\noindent are realizations from normal distributions centred at the
true (unknown) values and with standard deviations given by the
measurement uncertainties

 \begin{equation}
 \boldsymbol{\sigma}_{\mathbf{d}_{i}}=\left(\sigma_{m_{i}},\sigma_{\log P_{i}},\sigma_{\varpi_{i}},\sigma_{\left[\mathrm{Fe/H}\right]_{i}},\sigma A_{m_{i}}\right)\,. 
 \label{eq:unc-data}
 \end{equation}

{  Our test sample $\mathcal{D}=\left\{ \mathbf{d}_{i}\right\}
  _{i=1}^{N}$ consists of $N=200$ fundamental mode and first overtone
  RR Lyrae (RRL) stars with near-infrared (NIR) photometry
  ($\hat{m}_{i}={\hat{m}}_{K_{s}i}$) selected amongst the stars of the
  \cite{Dambis-et-al-2013} compilation for which TGAS parallaxes
  \citep{Lindegren-et-al-2016} and associated uncertainties were
  available.  This is essentially the same {$K_s$}-band sample of
  \citetalias{paperI} for which the periods of first overtone stars
  were "fundamentalised" by adding 0.127 to the decadic logarithm of
  the period and uncertainties on $\log\left(P\right)$ were estimated
  as $\sigma_{\log\left(P\right)}=0.01\cdot\log\left(P\right)$ which
  is equivalent to an uncertainty of 2\% in the period.  Unlike in
  \citetalias{paperI} (Sect. 6.1), where metal abundances were
  transformed from the \cite{Zinn-and-West-1984} to the
  \cite{Gratton-et-al-2004} metallicity scale to be consistent with a
  period term slope of the $PM_{K_{s}}Z$ relationship fixed to the
  value of $-2.73$ mag/dex reported by \cite{Muraveva2015}, in this
  paper we use the original metal abundances provided by
  \cite{Dambis-et-al-2013} because we aim to infer the period term
  slope. Because \cite{Dambis-et-al-2013} did not provide metallicity
  uncertainties, in \citetalias{paperI} we assigned a constant
  uncertainty of 0.2 dex to all metallicities in the sample.  In this
  paper we distinguish among techniques used to estimate metal
  abundances and respectively adopt uncertainties of 0.1, 0.2 and 0.3
  dex for metallicities estimated from high-resolution spectroscopy,
  measured by the $\Delta S$ method of \cite{Preston-1959} and
  determined from photometry or other non-spectroscopic methods. The
  \cite{Dambis-et-al-2013} catalogue does not include uncertainties on
  absorption. We estimate them as
  $\sigma_{A_{K}}=0.114\cdot\sigma_{A_{V}}$ , where
  $\sigma_{A_{V}}=3.1\cdot\sigma_{E\left(B-V\right)}$
  \citep{Cardelli-1989} and $\sigma_{E\left(B-V\right)}=0.16\cdot
  E\left(B-V\right)$ \citep{Schlegel-1998}. All measured quantities
  are represented as blue nodes in the DAG where we do not include the
  nodes corresponding to the uncertainties of Eq. \ref{eq:unc-data} in
  order to facilitate its interpretation.}

We represent the likelihood of the model with the nodes corresponding
to the true values $m_{i}$, $\varpi_{i}$, $\log P_{i}$,
${\left[\mathrm{Fe/H}\right]}_{i}$ and ${A_m}_i$ and the arcs going
from true values to measurements. { True values and observations}
are all enclosed in a black rectangle that represents replication for
the $N$ stars in the sample (plate notation). Equation
\ref{eq:plz-intro} can be written for every star $i$ in the sample as

\begin{equation}
  M_{i}=b+c\cdot\log P_{i}+k\cdot {\left[\mathrm{Fe/H}\right]}_{i}\,,
  \label{eq:plz1}
\end{equation}

\noindent where $M_i$ represents the true absolute magnitude for star
$i$.  This is a linear model in the parameters: the intercept $b$, the
slope $c$ for the period term, and the slope $k$ for the metallicity
term. The last term can be dropped if metallicities are thought to
play a negligible role in the relationship. We keep it in the
following for the sake of completeness, but the particularization to
PL relations is straightforward. In Fig. \ref{DAG} we shadow the left
hand rectangle that includes the metallicity terms to remark this
choice.

In Fig. \ref{DAG}, the $PL(Z)$ relationship of Eq. \ref{eq:plz1} is
denoted by the grey node $M_i$ and { all incoming arrows from $c$,
  $k$, $b$, $P_i$ and ${\left[\mathrm{Fe/H}\right]}_{i}$} (that is,
the three parameters and two predictive variables). In fact, the
reader may have noticed that there is an additional arrow linking $w$
and $M_i$. $w$ represents an intrinsic dispersion in the $PL(Z)$
relationship that may be due to evolutionary effects, for
example. This dependence on additional predictive variables that are
not accounted for in the model is incorporated as an additional
Gaussian spread of standard deviation $w$. This spread will be
analysed as part of the inference results. Including the additional
Gaussian spread that represents unaccounted predictive variables, {
  we have that}

\begin{equation}
M_{i}\sim\mathsf{N}\left(b+c\cdot\log
P_{i}+k\cdot\left[\mathrm{Fe/H}\right]_{i},w\right)\,,
\label{eq:plz-distr}
\end{equation}

\noindent { where} $\sim$ should be read as 'is distributed as',
$\mathsf{N}$ represents the normal (Gaussian) distribution, and the
comma separates values inside the parenthesis that represent the mean
and standard deviation of the normal distribution, respectively.

Of course, we do not observe absolute magnitudes, and our model has to
account for the transformation between absolute magnitudes and the
observations, that are (potentially affected by interstellar
absorption) apparent magnitudes. This is shown in the lower part of
Fig. \ref{DAG}, where the parallaxes (right-hand block) are handled as
we explain next.
   
The transformation from absolute to { unabsorbed} apparent
magnitudes is a well known deterministic one:

\begin{equation}
  {m_0}_i  =  M_{i} -5\log(\varpi_{i})+10\,,
  \label{eq:M-to-m0}
\end{equation}

\noindent where the parallax $\varpi_{i}$ is measured in mas. This is
not a probabilistic relation and we use dashed lines in the arrows
going into ${m_0}_i$ to distinguish them from the arcs denoting
conditional probability links. The { absorbed }apparent magnitudes
are computed as $m_{i}={m_0}_i+A_{m_{i}}$, where the gray node
${A_m}_i$ represents the true absorption. { The model also contemplates
the possibility of a TGAS global parallax offset $\varpi_{0}$. The
offset can be inferred by the model or fixed to a predefined
value. We shadow the top right rectangle of the graph that includes
the offset node to denote this choice.}

\subsection{Priors, hyperparameters and hyperpriors}
\label{sec:HM-Hyp} 
       
Prior distributions allow us to pose probabilistic statements about
plausible values of the model parameters based on knowledge available
prior to and independent of the observations. But most important, they
allow us by means of Bayes' theorem to make statements about the
distribution of the parameters we aim to infer (the posterior
distribution of the model parameters in the left side of Equation
\ref{eq:bayes2-1}). In the astrophysical context of this paper, we aim
at formulating probabilistic statements about the values of the
hyperparameters: the most probable value of the $PL(Z)$ slopes or
intercepts or their credible intervals. We use green rectangular nodes
at the top of the graph to denote fixed prior hyperparameters.

\begin{table*}[htb]
	\caption {Prior ($\pi$) definitions for the
		hierarchical Bayesian model of the $PL(Z)$
		relations. We use
		the $\pi$ symbol to refer to the prior probability.}
	\label{tab:priors}
	\centering
	\begin{tabular}{c}
		\hline
		\hline 
		\\	
		
		$\pi\left(M_{i}\mid b,c,k,w,\log P_{i},{[\rm{Fe/H}]}_{i}\right)=
		\mathsf{N}\left(b+ c\cdot\log\ P_{i} + k\cdot {[\rm{Fe/H}]}_{i},w\right)$ \\[1ex] 
		$\pi\left(b\right)=\mathsf{Cauchy}\left(0,10\right)$ \\[1.2ex]  
		$\pi\left(c\right)=\pi\left(k\right)=\mathsf{Cauchy}\left(0,1\right)$ \\[1ex]
		$\pi\left(w\right)=\mathsf{Exp}\left(1\right)$ \\[1ex] 
		$\pi\left(\left[\mathrm{Fe/H}\right]_{i},\log P_{i},\ln\varpi_{i}\mid\phi^{\left\{ k\right\} },
		\boldsymbol{\mu}^{\left\{ k\right\} },\mathrm{T}^{\left\{ k\right\} },\Omega^{\left\{ k\right\} }\right)=\sum_{k=1}^{2}\phi_{k}\mathsf{MN}\left(\boldsymbol{\mu}^{k},\mathrm{T}^{k}\Omega^{k}\mathrm{T}^{k}\right)$ \\[1ex] 
		$\pi\left(\boldsymbol{\phi}\right)=\mathsf{Dirichlet}\left(1,1\right)$ \\[1ex]
		$\pi\left(\boldsymbol{\mu}^{k}\right)=\mathsf{MN}\left(Q_{2k-1}\left(\left[\widehat{\mathrm{Fe/H}}\right],\log\hat{P},\ln\hat{\varpi}\right),\mathrm{diag}\left(0.2^{2},0.1^{2},0.05^{2}\right)\right)$ \\[1ex]
		$\pi\left(\sigma_{Z}^{k}\right)=\pi\left(\sigma_{P}^{k}\right)=\pi\left(\sigma_{\varpi}^{k}\right)=\mathsf{Exp}\left(1\right)$ \\[1ex]
		$\pi\left(\Omega_{P,\varpi}^{k}\right)=\mathsf{LKJ}\left(1\right)$ \\[1ex]
		$\pi\left(\varpi_0\right)=\mathsf{N}\left(0,0.1\right)$\\[1ex]

		\hline
	\end{tabular}
\end{table*}

{ Often the prior definitions used in the literature are
  conservative choices in the sense that they aim to be as
  non-informative as possible. For both slopes $c$ and $k$ of the
  $PL(Z)$ relationship of Eq. \ref{eq:plz1} we specify a standard
  Cauchy prior (centred at $0$ with scale parameter equal to $1$),
  which is equivalent to a uniform prior supported on the interval
  $\left[-\pi/2,+\pi/2\right]$ assigned to the angles
  $\theta_{1}=\arctan\left(c\right)$ and
  $\theta_{2}=\arctan\left(k\right)$.  The prior probability
  distribution of the intercept is a Cauchy distribution centred at
  its mode $\mu_{b}=0$ with scale parameter $\sigma_b=10$. The
  intrinsic dispersion of the $PLZ$ relation is given by an exponential
  prior with inverse scale $\lambda_w=1$. }
   
{ The only block of the graph that remains to be clarified is the
  left-hand top block describing the distribution of periods,
  metallicities and parallaxes. Our model assigns a joint
  three-dimensional prior to the true distribution of metallicities,
  periods and parallaxes. Furthermore, we distinguish two components
  with different chemical composition and a different relationship
  between pulsation period and parallax (distance). The prior is
  defined as a mixture of Gaussian distributions (Gaussian mixture, GM) given by}

\begin{equation}
\left(\left[\mathrm{Fe/H}\right]_{i},\log P_{i},\ln\varpi_{i}\right)\sim\sum_{k=1}^{2}\phi_{k}
\mathsf{MN}\left(\boldsymbol{\mu}^{k},\mathrm{T}^{k}\Omega^{k}\mathrm{T}^{k}\right)\,,
\label{eq:P-Z-Pi-dist}
\end{equation}

{ \noindent where $\phi_{k}$ represents the mixing proportion of the
  $k$-th component of the mixture, and $\mathsf{MN}$ is a 3D Gaussian
  probability density with mean vector
  $\boldsymbol{\mu}^{k}=(\mu_{Z}^{k},\mu_{P}^{k},\mu_{\varpi}^{k})$,
  diagonal matrix of standard deviations
  $\mathrm{T}^{k}=\mathrm{diag}(\sigma_{Z}^{k},\sigma_{P}^{k},\sigma_{\varpi}^{k})$
  and correlation matrix }

\begin{equation}
\Omega^{k}=\left(\begin{array}{cc}
1 & \begin{array}{cc}
0 & 0\end{array}\\
\begin{array}{c}
0\\
0
\end{array} & \Omega_{P,\varpi}^{k}
\end{array}\right)\,.
\label{eq:test}
\end{equation}

{ The parameters of the prior defined by Equation
  \ref{eq:P-Z-Pi-dist} are themselves model parameters and subject of
  the Bayesian inference as well.  As such, they have their own
  hyperpriors. We assign a Gaussian prior centred at the first ($Q_1$)
  and third ($Q_3$) quartiles of the distribution of the measurements
  $\left(\left[\widehat{\mathrm{Fe/H}}\right],\log\hat{P},\ln\hat{\varpi}\right)$
  and a covariance matrix equal to
  $\mathrm{diag}\left(0.2^{2},0.1^{2},0.05^{2}\right)$ for the mean
  vector $\boldsymbol{\mu}^{k}$ of each mixture component. This prior
  is chosen to prevent the non-identifiability of mixture components
  (the posterior multimodality arising from the fact that swapping the
  label of the two 3D Gaussian components results in exactly the same
  solution). We assign a weakly informative exponential prior with
  inverse scale $\lambda_{\mathrm{T}^k}=1$ to the standard deviations
  in each $\mathrm{T^k}$.  For each correlation submatrix
  $\Omega_{P,\varpi}^{k}$ we specify a LKJ prior
  \citep{Lewandowski-2009} with $\nu=1$ degrees of freedom. The choice
  of prior given by Equation \ref{eq:P-Z-Pi-dist} will prove to be
  critical for the correct inference of the PLZ relation coefficients
  for reasons that will become apparent in Sect.
  \ref{sec:validation}.}

Fig. \ref{DAG} translates into the following likelihood
\begin{equation}
  \begin{split}
    p\left(\mathcal{D}\mid\mathbf{\Theta}\right)=
    &\prod_{i=1}^{N}p\left(\mathbf{d}_{i}\mid\mathbf{\Theta}\right)
    = \prod_{i=1}^{N}p\left(\widehat{[\rm{Fe/H}]}_{i}\mid {[\rm{Fe/H}]}_{i},\sigma_{Z_{i}}\right) \\
    & \hphantom{=\prod_{i=1}^{N}} \cdot p\left(\log\hat{P}_{i}
    \mid\log P_{i},\sigma_{P_{i}}\right) \\ 
    & \hphantom{=\prod_{i=1}^{N}} \cdot  p\left(\hat{m}_{i}\mid M_{i}
    ,A_{m_{i}},\varpi_{i},\sigma_{m_{i}}\right) \\
    & \hphantom{=\prod_{i=1}^{N}} \cdot p\left(\hat{\varpi}_{i}\mid\varpi_{i},
    \sigma_{\varpi_{i}},\varpi_0\right) \\
    & \hphantom{=\prod_{i=1}^{N}} \cdot p\left(\hat{A}_{m_{i}}\mid A_{m_{i}},\sigma_{A_{m_i}}\right)  
  \end{split}
\end{equation}

\noindent and priors:
   
\begin{equation}
  \begin{split}
    \pi\left(\mathbf{\Theta}\right) 
    &= \pi\left(\phi^{\left\{ k\right\} },\boldsymbol{\mu}^{\left\{ k\right\} },\mathrm{T}^{\left\{ k\right\} },\Omega_{P,\varpi}^{\left\{ k\right\} }\right)    \cdot\pi\left(b,c,k,w\right)\cdot\pi\left(\varpi_0\right)
     \\ 
    & \hphantom{=} \cdot\prod_{i=1}^{N}  \pi\left(\left[\mathrm{Fe/H}\right]_{i},\log P_{i},\ln\varpi_{i}\mid\phi^{\left\{ k\right\} },\boldsymbol{\mu}^{\left\{ k\right\} },\mathrm{T}^{\left\{ k\right\} },\Omega^{\left\{ k\right\} }\right)\\
    & \hphantom{=\prod_{i=1}^{N}} \cdot\pi\left(M_{i}\mid b,
    c,k,w,\log P_{i},{[\rm{Fe/H}]}_{i}\right) \\ 
  \end{split}
\end{equation}

\noindent where each prior probability is defined in Table
\ref{tab:priors}.
 
We have encoded our HM using the Stan probabilistic modelling language
\citep{Carpenter-et-al-17} and used the No-U-Turn sampler (NUTS) of
\cite{Hoffman-Gelman-14} to compute the MCMC samples corresponding to
the parameters of interests.
   
\section{Model validation with semi-synthetic data.}
\label{sec:validation}    

{ In this section we aim at validating the improved HM described in
  Sect. \ref{sec:hiermodel} on synthetic data as close as possible to
  true data sets but following exactly an RR Lyrae $PM_{K_s}Z$
  relation from the literature.} We simulate { three sets A, B and
  C} of semisynthetic true absolute magnitudes and parallaxes using
that $PM_{K_s}Z$ relation and the apparent magnitudes of the sample
described in Sect. \ref{sec:hiermodel}.  The only difference between
the synthetic data sets A and B lies in the assumed parallax
uncertainties. In data set A we generate parallax uncertainties from
an hypothesized distribution, and in data set B we use the TGAS
uncertainties. { No parallax offset is introduced in the
  simulations. The data set C is identical to data set A except for
  the uncertainty on metallicity measurements that is reduced by
  half}. Our objective is to analyse the impact of the hyperprior
choice, the influence of the parallax { and metallicity}
uncertainties on the inferred coefficients under these { three}
scenarios and detect potential biases in the sample.  

\subsection{Validation data sets}
\label{sec:semi-synth-samples} 
 
{ In what follows, we describe the construction of the three data
  sets A, B and C that reproduce the generative process of the
  observations of Eq. \ref{eq:data} from the model hyper-parameters
  for the three simulated scenarios.  In the generation of the
  semi-synthetic samples we first used a Bayesian model to draw
  individual true metallicities and logarithms of the true periods
  from Gaussian posterior distributions inferred from their
  measurements and associated uncertainties in the real data set
  described in Sect. \ref{sec:HM-Hyp} with vague Gaussian priors
  assigned to both sets of parameters. Then the observed values were
  drawn from a Gaussian distribution centred at the true value and
  with standard deviation given by the measurement uncertainties. For
  the particular case of data set C the metallicity measurements have
  been generated dividing by 2 the uncertainties in the real
  sample. We generated true absolute magnitudes from the $PM_KZ$
  theoretical calibration of \cite{Catelan-2004} adopting
  $\left[\alpha/\mathrm{Fe}\right]=0.3$ and converting from
  $\left[\mathrm{Fe/H}\right]$ to $\log Z$ by means of its Eqs. (9)
  and (10):}

\begin{equation}
  M_{K_{s}}=-2.353\log P+0.175\left[\mathrm{Fe/H}\right]-0.869.
  \label{eq:PKZ-Catelan-2004}
\end{equation}

We generated true parallaxes from

\begin{equation}
  \varpi_{i}=10^{2+0.2\left(M_{i}-{m_0}_{i}\right)}\,,
  \label{eq:sim-true-phi}
\end{equation}
   
\noindent where ${m_0}_i$ is the unabsorbed apparent magnitude of the
$i$-th star in the real sample. For data set A we generated measured
parallaxes from a Gaussian distribution centred at the true parallaxes
given by Eq. \ref{eq:sim-true-phi} with standard deviations drawn from
an exponential distribution with inverse scale parameter equal to $10$
plus a zero-point of 0.01 mas. { We note that the uncertainties on
  TGAS parallaxes are higher than these ones by approximately one
  order of magnitude, but our objective here is to evaluate the
  performance of our HM under the small uncertainties typical of the
  \textit{Gaia} DR2.} The measured parallaxes of data set B were
generated using the TGAS parallax uncertainties.

\begin{figure}
  \begin{center}
    \hspace*{-0.01cm}
    \includegraphics[scale=0.5]{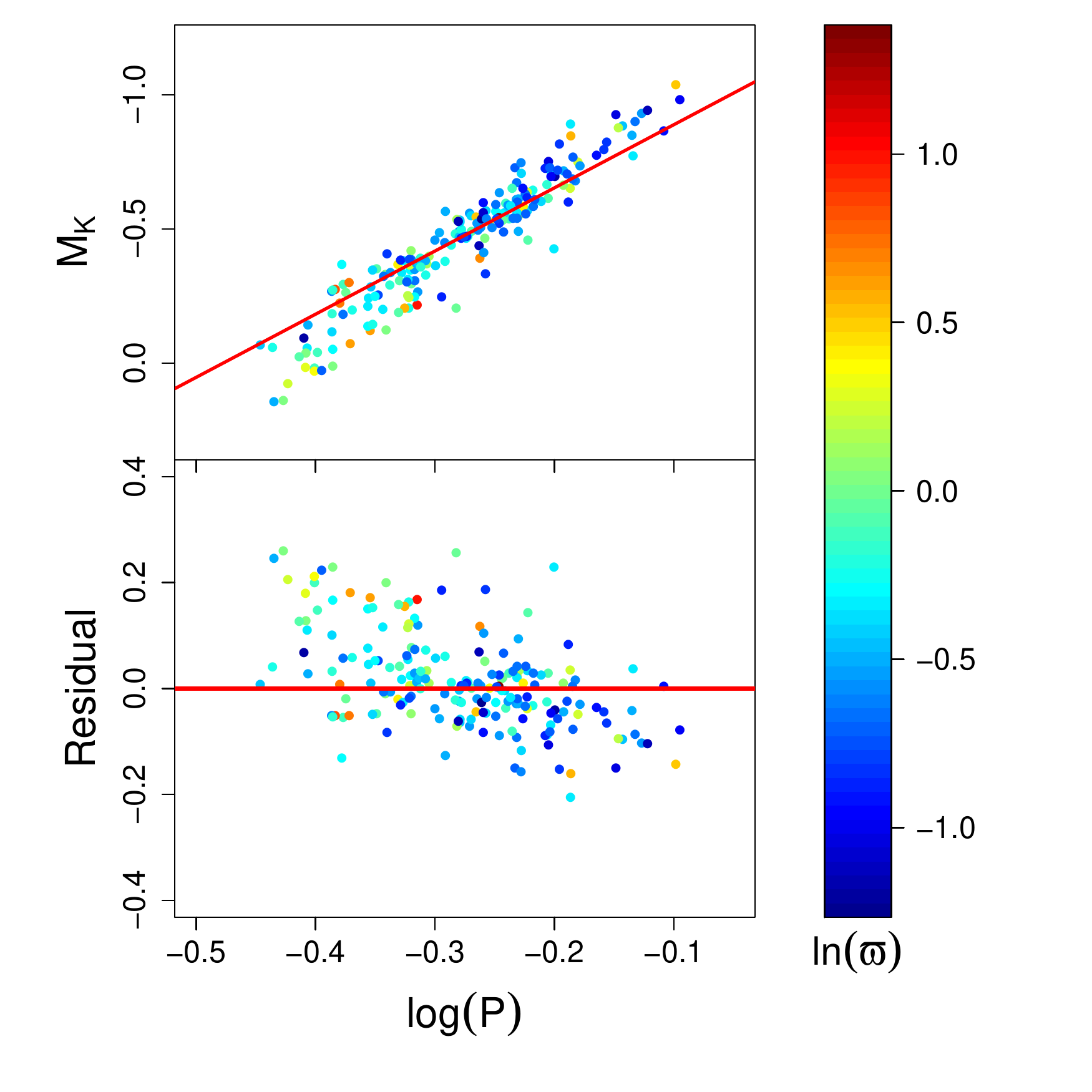}
   		
    \caption{Plot of the observed absolute magnitudes as a function of
      the true decadic logarithm of the period for simulations A and B
      The red solid line represents the projection of the $PM_{K_s}Z$
      relation of Eq.  \ref{eq:PKZ-Catelan-2004} for a value of the
      true metallicity equal to the median of the values generated
      according to the text. Colours encode the simulated true
      parallaxes according to the logarithmic scale on the right.}
	\label{fig:sims-A-B-PL}
    \end{center}
    
\end{figure}
 
\begin{figure}
  \begin{center}
    \includegraphics[scale=0.50]{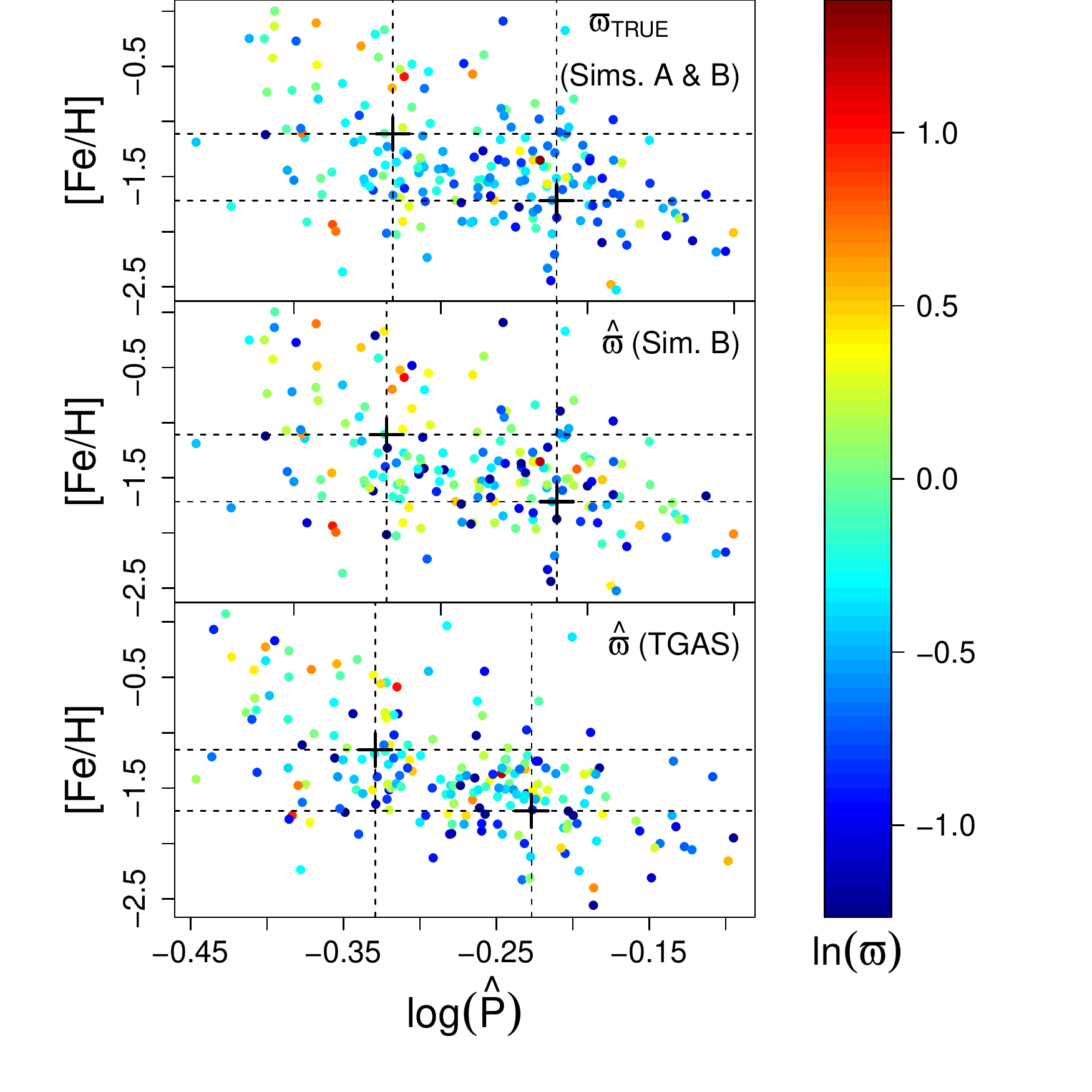}
    
    \caption{Scatter plot of observed metallicities and $\log(P)$ for
      validation sets A and B and the real sample used in the
      paper. The black crosses represent the first ($Q_1$) and third
      ($Q_3$) quartile of the distribution of measured $\log(P)$ and
      ${[\rm{Fe/H}]}$ for simulations A and B (top and middle panel)
      and the real sample (bottom panel). The colour encodes the
      natural logarithm of true parallaxes of simulations A and B (top
      panel), measured parallaxes of simulation B (middle panel) and
      TGAS parallax estimates (bottom panel).  }
    \label{fig:corr-met-logp}
    \end{center}
\end{figure}

{ Figure \ref{fig:sims-A-B-PL} represents in the top panel the
  values of the absolute magnitudes and logarithms of the true periods
  generated for simulations A and B. The red line shows the PL
  relation for the median value of $[\rm{Fe/H}]$ in the simulated
  sample. Because this plot is a 2D projection of the 3D $PM_{K_s}Z$
  relation and the absolute magnitudes were sampled from it, any
  deviations from the red line can only be explained by metallicities
  differing from the median and the intrinsic dispersion of the
  relation (which is symmetric).  A correlation between periods and
  metallicities is evident which results in correlated residuals
  (lower panel) with respect to the assumed PL for the median true
  metallicity.  We also observe in the figure a correlation with the
  distance: brighter magnitudes correspond (on average) to longer
  periods and lower metal abundances at larger distances (lower
  parallaxes) and {\sl viceversa}. Figure \ref{fig:corr-met-logp}
  demonstrates that the correlation between periods, metallicities and
  parallaxes is not an added effect in the simulations and is present
  in the measurements both for the simulated and the real data
  set. The figure represents the measured metallicities vs.\ the
  logarithm of the period for simulations A and B and for the real
  sample. The colour code reflects the natural logarithm of the
  parallaxes: the same true value for simulations A and B in the top
  panel, the observed value for simulation B in the middle panel, and
  the measured value of the TGAS catalogue in the bottom panel. The
  black crosses denote the first and third quartiles of the marginal
  distributions along each axis. In principle, one expects the
distribution of periods to be independent of distance. However,
Fig. \ref{fig:corr-met-logp} shows that the left half of the plot
(stars with short periods) is predominantly populated by stars with
larger parallaxes and higher metallicities while the right half is, again on
average, predominantly populated by distant stars (smaller parallaxes)
with lower metallicities. We observe that the correlation between the
simulated true parallaxes, measured periods and metallicities shown by
the top panel of the figure persists for the measured parallaxes
depicted in the middle panel, although with larger dispersion due to
the higher parallax uncertainties of simulation B. We also note that
the correlation is also present in the real sample of TGAS parallaxes
(bottom panel). The interpretation is as follows: we expect nearby
stars in our sample to be characterised on average by the higher
metallicities of the disk, while the opposite is true for those
further away in the halo. This dependence of distance on metallicity
is visible in the colour code of Fig. \ref{fig:corr-met-logp} and
crudely characterized by the two black crosses on each panel. Each cross
represents the projection of the point
$Q_{k}\left(\left[\mathrm{Fe/H}\right],\log P,\ln\omega\right)$ (with
$Q_k$ denoting the $k$-th quartile for $k=1,2$ ) onto the
period-metallicity plane. The crosses on the top panel of the figure
correspond respectively to distances of $1.87$ and $1.07$ kpc for a
lower ($-1.72$ dex) and higher ($-1.11$ dex) metallicity component.}

{ The correlations just described show themselves on the period
  luminosity diagram of Fig.  \ref{fig:sims-A-B-PL} in the following
  way.} Because higher metallicities correspond to shorter periods (as
illustrated in Fig. \ref{fig:corr-met-logp}) we then expect the nearby
stars (that let us recall, are on average more metal-rich) to be
characterised by shorter periods (the left half of the PL
diagram). And the opposite is also true: the distant (small parallax)
halo stars have on average lower metallicities and hence, { longer
periods} (the right half of the PL diagram).  This scenario is then
prone to systematic biases in the $\log(P)$ slope inference results
because it is precisely at the right edge of the PL diagram that there
is a concentration of the most distant sources that will inevitably be
characterised by larger fractional parallax uncertainties. We know
that in general, the prior plays a minor role whenever the
uncertainties are small because a narrow likelihood dominates the
posterior. And the opposite is true for stars with large fractional
parallax uncertainties: the likelihood is barely informative and it is
the prior that dominates the posterior. \cite{Bailer-Jones-2015}
shows very pedagogic illustrations of this for the problem of
inferring distances for parallaxes under several prior
specifications. In our case, the stars for which the prior has a
larger impact on the inference of the parallax are predominantly
placed at the rightmost range of periods.

\begin{figure}
	\begin{center}
		\includegraphics[scale=0.55]{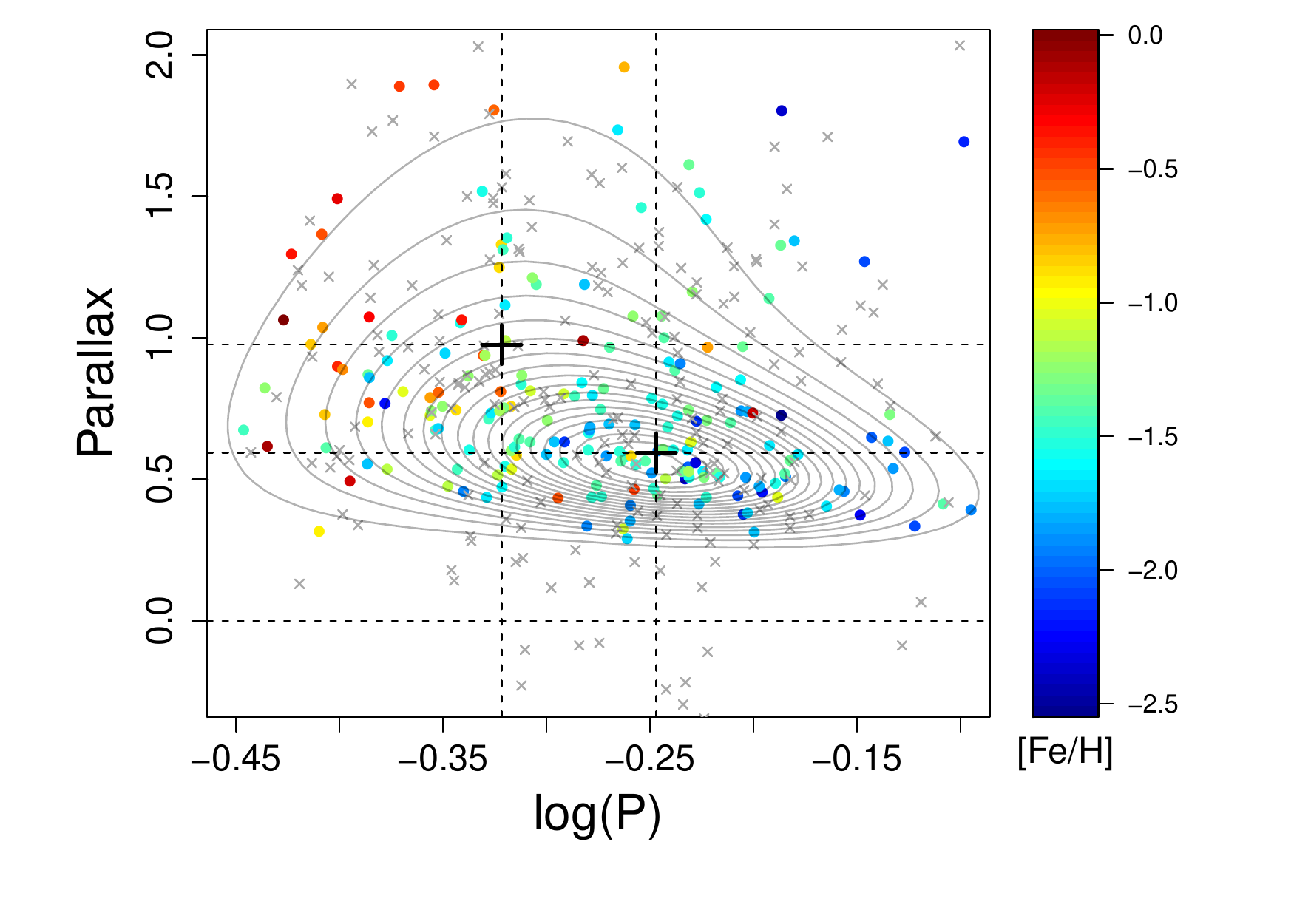}
		\caption{Projection of a two components 3D GM PDF 
			fitted to the true values of 
			$\left(\left[\mathrm{Fe/H}\right],\log P,\ln\varpi \right)$ 
			in the data set B onto the plane period-parallax. 
			The gray contours represent iso-probability lines.
			The coloured and gray points represent respectively 
			true and measured parallaxes. The black crosses indicate 
			the median of the mean posterior distribution of 
			each GM component. The colour encodes the metallicity. }
		\label{fig:3D-log-prior-projection}
	\end{center}
\end{figure}

These relatively hidden correlations will have important consequences
for the inference as we will see. In particular, it will have an
impact on the choice of prior. In general, the parallax prior has to
have support (non-vanishing values) in all the range of true
parallaxes. But this is even more important given the correlation
between periods and true parallaxes because in the case of our
simulations, long periods have on average the smallest parallaxes.
{ If the prior has zero probability density for the small true
  parallaxes, and given the relatively large parallax uncertainties in
  our sample, the model will systematically assign parallaxes larger
  than the true ones (will overestimate them).  Given that in both
  simulated data sets, the full range of absolute magnitudes is
  reduced to a brighter magnitude range for lower parallaxes, and
  because the strong deterministic relationship between absolute
  magnitudes and true parallaxes established by Eq. \ref{eq:M-to-m0}
  of the HM, the model will infer absolute magnitudes fainter than the
  true (brighter) ones.  Finally, if the stars with long periods get
  fainter absolute magnitudes, the model will systematically
  underestimate the absolute value of the period slope coefficient of
  the $PLZ$ relation. Hence, if our interpretation is correct, the
  distance prior has to be made dependent on metallicity. The model
  described in Sect. \ref{sec:hiermodel} addresses this problem by
  using a 3D prior that distinguishes between two probabilistic
  classes of metal abundance in the data and constrains the true
  parallaxes of each class by means of their relationship with
  periods. } { Figure \ref{fig:3D-log-prior-projection} shows an
  example of our 3D GM prior fitted to the distribution of true
  $\left(\left[\mathrm{Fe/H}\right],\log P,\ln\varpi \right)$ in
  simulation B using a Bayesian HM. The figure represents the
  projection of the fitted probability density function (PDF) onto 
  the plane period-parallax (with
  the parallax represented in linear scale) and depicts the measured
  pairs $\log(P)$-parallax in the simulated dataset.} { We note
  that without a proper modelling of the selection effect that gives
  rise to the correlation between periods and true parallaxes, the
  inference will return a severely underestimated $\log(P)$ slope, as
  we will demonstrate towards the end of this section, where our model
  will be compared with a model that uses 1D independent prior
  distributions for period, metallicity and parallax.  This comparison
  will illustrate the shrinkage power of hierarchical models.}


\subsection{Validation results}
\label{sec:valid-results} 

\begin{figure}
  \begin{center}
    \includegraphics[scale=0.5]{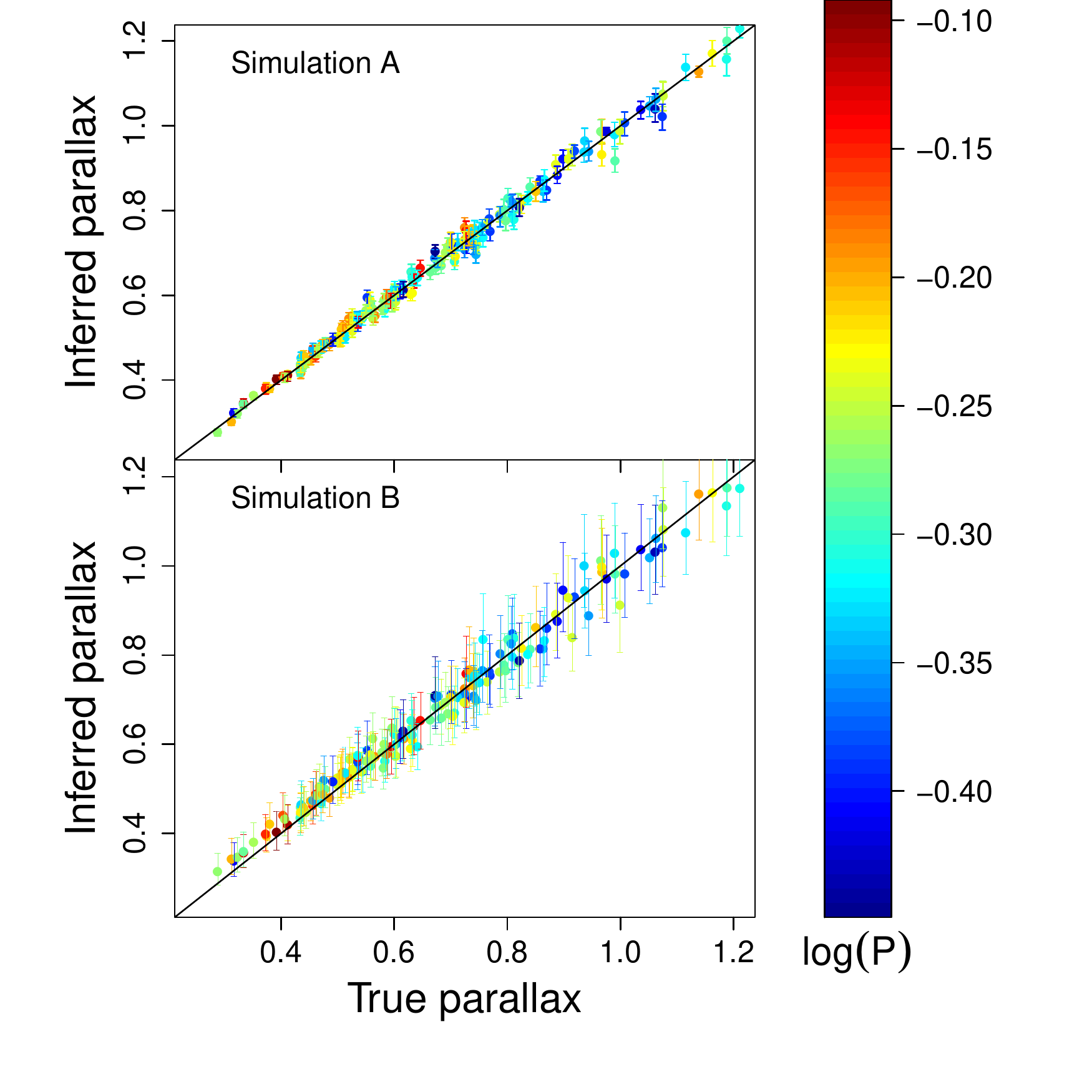}
    
    \caption{Comparison between inferred and true parallaxes for
      validation sets A and B. The solid lines represent the bisectors
      and the colour encodes the value of the decadic logarithm of the
      period in days according to the colour scale on the right.}

    \label{fig:inf-pi-sim-pi}
  \end{center}
\end{figure}

{ Figure \ref{fig:inf-pi-sim-pi} compares the parallaxes inferred
  by our HM with the true ones simulated for datasets A (top panel)
  and B (bottom panel). The colours in both panels represent the
  $\log(P)$ according to the colour scale to the right. For simulation
  B we observe that lower parallaxes (typical of the longer periods)
  are slightly overestimated. We also observe that the smallest (and
  overestimated) parallaxes correlate as expected with longer
  periods. But the largest parallax overestimations do not correspond
  systematically to the longest periods which is guaranteed by the 3D
  prior used in our HM. The overestimation of the smallest parallaxes
  is interpreted as a deficiency of our Gaussian 3D prior that fails
  to represent adequately the more complex spatial distribution of
  RR~Lyrae stars in the Galaxy.}

\begin{figure}
  \begin{center}
    \includegraphics[scale=0.5]{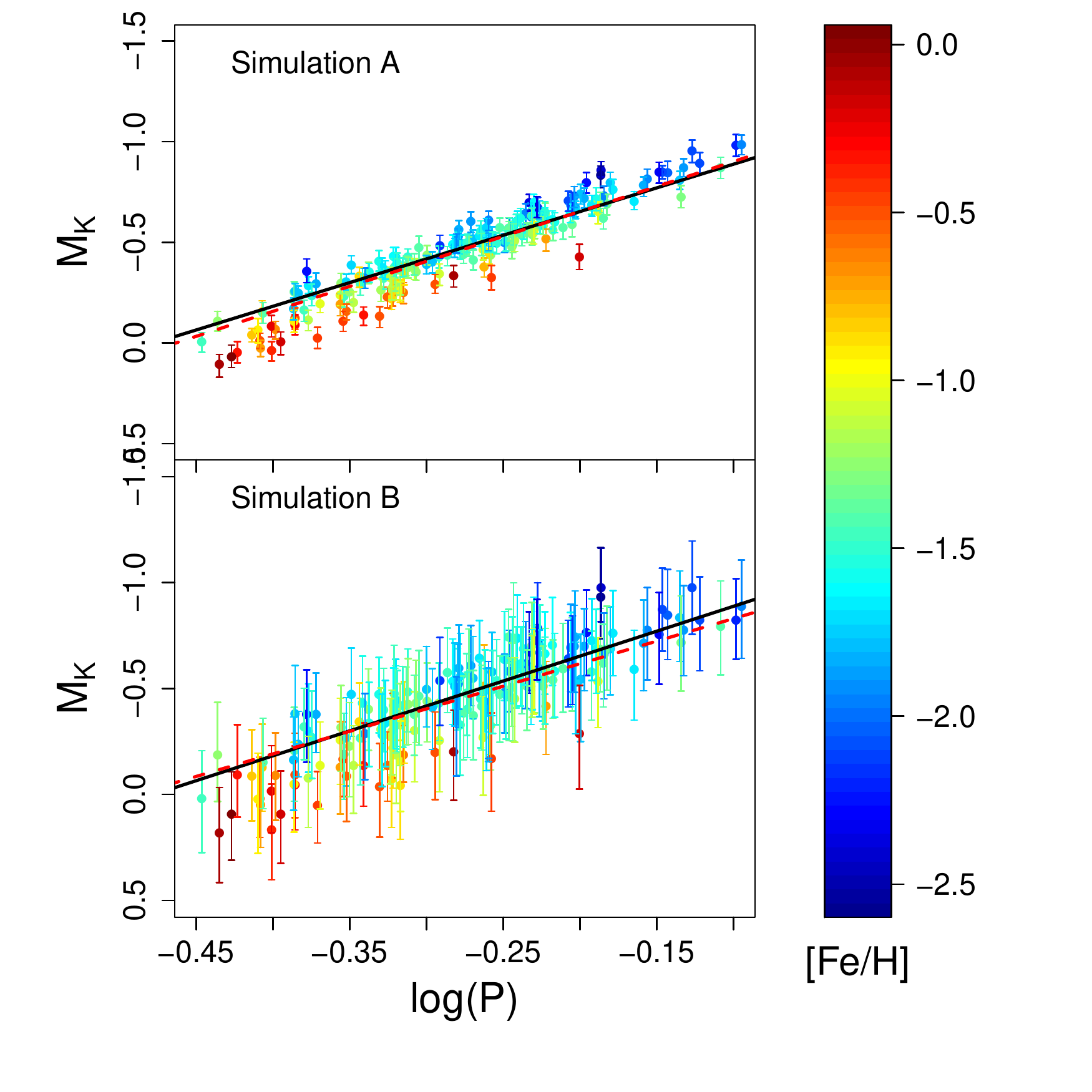}

    \caption{Comparison between inferred absolute magnitudes and true
      periods for validation sets A and B.  The black-solid and
      red-dashed lines represent respectively the projections of the
      $PM_KZ$ relation of Eq.  \ref{eq:PKZ-Catelan-2004} and the
      relation inferred by our HM (adopting the median value of the
      posterior distribution of each coefficient) for a value of the
      true metallicity equal to the median of the values generated
      according to the text.  The colour encodes the metallicity
      according to the scale on the right.}

    \label{fig:inf-sim-M-P}
  \end{center}   	
\end{figure}

{ Figure \ref{fig:inf-sim-M-P} compares, for simulations A and B,
  the inferred absolute magnitudes with the true periods and
  represents the $PM_KZ$ relation inferred in each case. Table
  \ref{tab:sim-fitting-comp} summarises the coefficients of the
  resulting $PM_KZ$ relations providing 68\% credible intervals around
  the median of the posterior distributions of them.  For simulation B
  (bottom panel of Fig.\ref{fig:inf-sim-M-P}) we observe that brighter
  magnitudes are slightly underestimated for long period stars.  This
  mild bias is not present for the data set A (upper panel) because
  the negligible parallax uncertainties tightly constrain the model
  parameters (the true parallaxes and hence, the absolute magnitudes
  and the slopes of the relation).  To better appreciate this results
  we suggest the reader also compare the inferred absolute magnitudes
  of the two scenarios of Fig. \ref{fig:inf-sim-M-P} with their
  simulated values on Fig. \ref{fig:sims-A-B-PL}.  The mild
  underestimation of brighter magnitudes for simulation B translated
  into a mild underestimation of the inferred $\log(P)$ slope (second
  row of Table \ref{tab:sim-fitting-comp}) with a credible interval of
  ${-2.13}_{-0.78}^{+0.75}$ mag/dex which in any case is in good
  agreement with the value $c=-2.35$ used for the simulation taken
  into account the large parallax uncertainties in this case. For
  simulation A the credible interval obtained for the $\log(P)$ slope
  was ${-2.48}_{-0.14}^{+0.15}$ mag/dex which slightly overestimate
  $c=-2.35$.  We hypothesize that this mild overestimation of the
  $\log(P)$ slope is a consequence of the correlation between period
  and metallicity and the relatively large uncertainties of measured
  metallicities. In order to evaluate this hypothesis we have used the
  third semi-synthetic data set (labelled C) whose uncertainties on
  simulated metallicity measurements are of the order of magnitude
  typical for high-resolution spectroscopic techniques.  The results
  for simulation C are included in the bottom row of Table
  \ref{tab:sim-fitting-comp} and seem to confirm our intuition.}

\begin{table*}

  \caption {Coefficients of the $PM_KZ$ relations inferred from the
    measurements of simulations A, B and C: slopes ($c$ and $k$), 
    zero-point ($b$) and intrinsic dispersion ($w$).  The posterior
    distribution of coefficients is summarized by the median plus
    minus the difference in absolute value between the median and the
    84th and 16th percentile (first line) and the maximum a posteriori
    (MAP) estimate (second line) for each simulation.  }
  \label{tab:sim-fitting-comp}

  \centering 
  \begin{tabular}{lllll}
    \hline
    \hline
 
    Simulation & \multicolumn{1}{c}{$c$} & \multicolumn{1}{c}{$k$} & 
    \multicolumn{1}{c}{$b$} & \multicolumn{1}{c}{$w$}\\
      & \multicolumn{1}{c}{(mag/dex)}   & \multicolumn{1}{c}{(mag/dex)}  & \multicolumn{1}{c}{(mag)}    & \multicolumn{1}{c} {(mag)}  \\

    \hline
    A & ${-2.48}_{-0.14}^{+0.15}$ &  ${0.15}_{-0.02}^{+0.03}$
    & ${-0.93}_{-0.06}^{+0.07}$ & ${0.03}_{-0.01}^{+0.02}$ \\
    & $-2.51$ & $0.15$ & $-0.94$ & $0.03$\\
    B & ${-2.13}_{-0.78}^{+0.75}$ &  ${0.22}_{-0.10}^{+0.10}$ 
    & ${-0.72}_{-0.31}^{+0.31}$  & ${0.20}_{-0.07}^{+0.07}$ \\
    & $-2.03$ & $0.22$ & $-0.71$ & $0.21$\\		
    C & ${-2.39}_{-0.14}^{+0.15}$ &  ${0.17}_{-0.03}^{+0.02}$ 
    & ${-0.89}_{-0.07}^{+0.07}$  & ${0.03}_{-0.071}^{+0.02}$ \\
    & $-2.39$ & $0.17$ & $-0.88$ & $0.03$\\
    \hline

  \end{tabular}	
\end{table*}

  \begin{table*}[htb]
  	
  	\caption {Coefficients of the $PM_KZ$ relations inferred for the
  		simulated scenarios A and B by an HM using the 3D GM prior and
  		two 1D parallax priors (the log-normal prior used in
  		\citetalias{paperI} and the EDVD prior of
  		\cite{Bailer-Jones-2015}): slopes ($c$ and $k$), zero-point
  		($b$) and intrinsic dispersion ($w$).  The posterior
  		distribution of each coefficient is summarized as in Table
  		\ref{tab:sim-fitting-comp}.  }
  	
  	\label{tab:sims-priors-MCMC-statistics}
  	
  	\centering 
  	\begin{tabular}{llllll}
  		\hline
  		\hline
  		
  		Simulation & Prior & \multicolumn{1}{c}{$c$} & \multicolumn{1}{c}{$k$} & 
  		\multicolumn{1}{c}{$b$} & \multicolumn{1}{c}{$w$}\\
  		&  & \multicolumn{1}{c}{(mag/dex)}   & \multicolumn{1}{c}{(mag/dex)}  & \multicolumn{1}{c}{(mag)}    & \multicolumn{1}{c} {(mag)}  \\

  		\hline
  		& 3D GM & ${-2.48}_{-0.14}^{+0.15}$ &  ${0.15}_{-0.02}^{+0.03}$
  		& ${-0.93}_{-0.06}^{+0.07}$ & ${0.03}_{-0.01}^{+0.02}$ \\		
  		&	& $-2.51$ & $0.15$ & $-0.94$ & $0.03$ \\
  		
  		A  & EDVD & ${-2.46}_{-0.15}^{+0.16}$ &  ${0.16}_{-0.03}^{+0.03}$ 
  		& ${-0.92}_{-0.07}^{+0.08}$  & ${0.03}_{-0.01}^{+0.02}$ \\	
  		&	& $-2.48$ & $0.15$ & $-0.92$ & $0.03$\\

  		& Log-Normal & ${-2.46}_{-0.15}^{+0.16}$ &  ${0.15}_{-0.03}^{+0.03}$ 
  		& ${-0.93}_{-0.07}^{+0.07}$  & ${0.04}_{-0.01}^{+0.02}$ \\	
  		&	& $-2.47$ & $0.15$ & $-0.94$ & $0.03$\\

  		\hline
  		& 3D GM & ${-2.13}_{-0.78}^{+0.75}$ &  ${0.22}_{-0.10}^{+0.10}$ 
  		& ${-0.72}_{-0.31}^{+0.31}$  & ${0.20}_{-0.07}^{+0.07}$ \\		
  		&  & $-2.03$ & $0.22$ & $-0.71$ & $0.21$\\
  		
  		B & EDVD & ${-1.31}_{-0.67}^{+0.63}$ &  ${0.12}_{-0.10}^{+0.10}$ 
  		& ${-0.62}_{-0.29}^{+0.28}$  & ${0.18}_{-0.06}^{+0.07}$ \\		
  		&	& $-1.32$ & $0.12$ & $-0.62$ & $0.17$\\

  		& Log-Normal & ${-0.95}_{-0.63}^{+0.59}$ &  ${0.08}_{-0.10}^{+0.10}$ 
  		& ${-0.52}_{-0.27}^{+0.26}$  & ${0.24}_{-0.06}^{+0.06}$ \\		
  		&	& $-0.87$ & $0.10$ & $-0.50$ & $0.25$\\
  		
  		\hline
  	\end{tabular}	
  \end{table*}

\begin{figure}[]
	\begin{center}
		\includegraphics[scale=0.5]{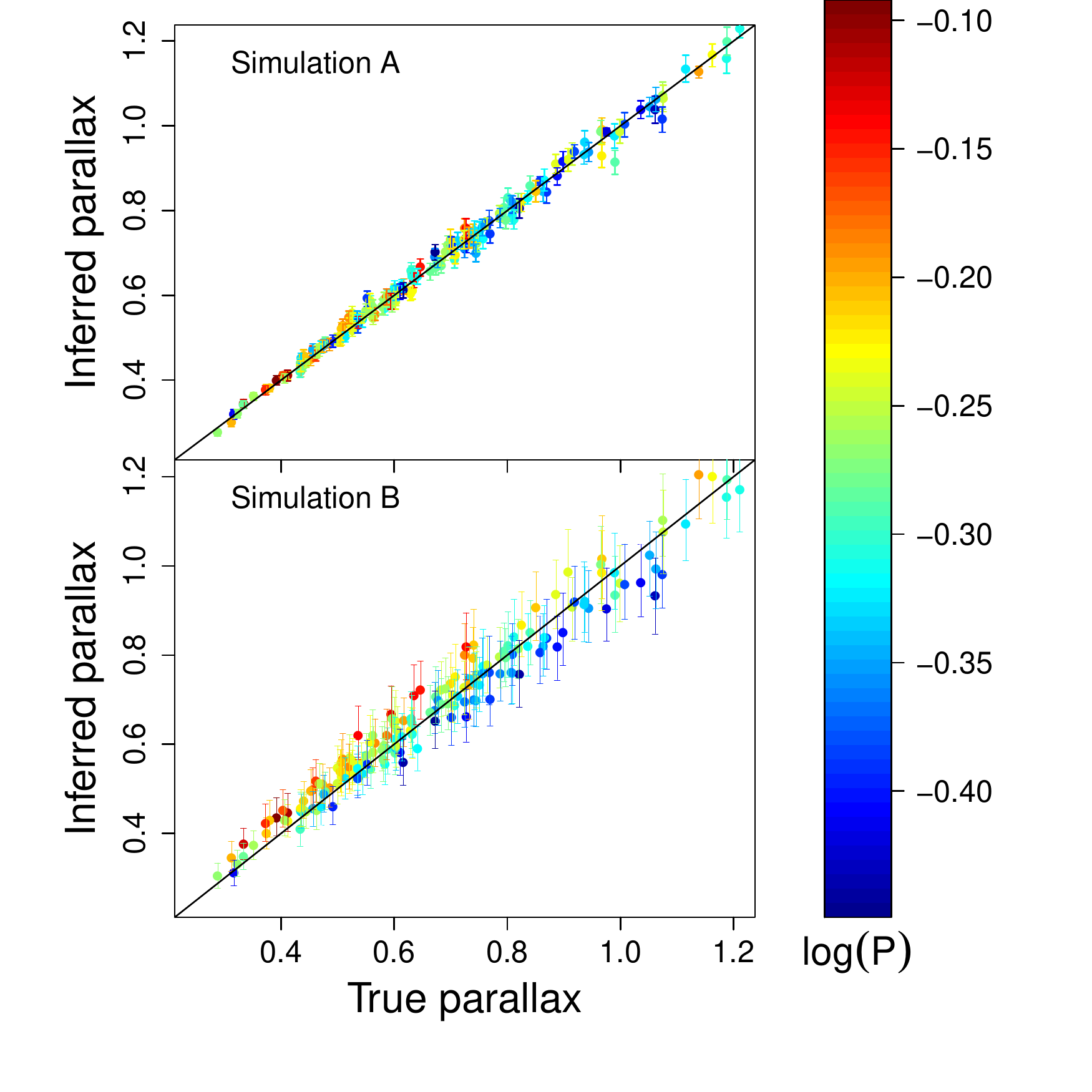}
		
		\caption{Comparison between inferred and true
                  parallaxes for validation sets A and B when a HM
                  with an EDVD prior is used. The colour encodes the
                  value of the decadic logarithm of the period in days
                  according to the colour scale on the right.}
		
		\label{fig:inf-pi-sim-pi-EDVD}
	\end{center}
\end{figure}

{ Table \ref{tab:sims-priors-MCMC-statistics} compares the results
  obtained with the 3D GM prior discussed above with those of
  alternative 1D priors based on the log-normal prior used in
  \citetalias{paperI} and the EDVD prior of \cite{Bailer-Jones-2015}
  given by Equation \ref{eq:EDVD-prior}.

\begin{equation}
    p\left(\varpi\right)=\frac{1}{2L^{3}\varpi^{4}}\exp\left(-\frac{1}{\varpi L}\right)\,.
	\label{eq:EDVD-prior}
\end{equation}

It shows that the existing correlations amongst periods, parallaxes
and metallicities have a small impact in the context of the small
parallax uncertainties that characterise simulation A, but affect
(worsen) significantly the inference outcome for the typical TGAS
uncertainties.} { Figure \ref{fig:inf-pi-sim-pi-EDVD} illustrates
  these facts by comparing the parallaxes inferred by an HM with an
  EDVD prior with the true parallaxes of both semi-synthetic data sets
  A and B.  We observe that in simulation B (bottom panel) the
  overestimation and underestimation of the inferred parallaxes is
  severe as compared to Fig. \ref{fig:inf-pi-sim-pi} of
  Sect. \ref{sec:validation}.}

  { It is important to bear in mind that although succesive
    \textit{Gaia} data releases will tend to decrease the measurement
    uncertainties in general, the community will often work with
    samples of stars (not necessarily classical pulsators like
    in the case of \citetalias{paperI}; \citealt{paperIII})
    with \textit{Gaia} uncertainties in the range exemplified by 
    our TGAS sample. }

\section{Application to the RR Lyrae  Gaia DR1 data } 
\label{sec:post}
   
\begin{figure*}
  \begin{center}
    \includegraphics[scale=0.4]{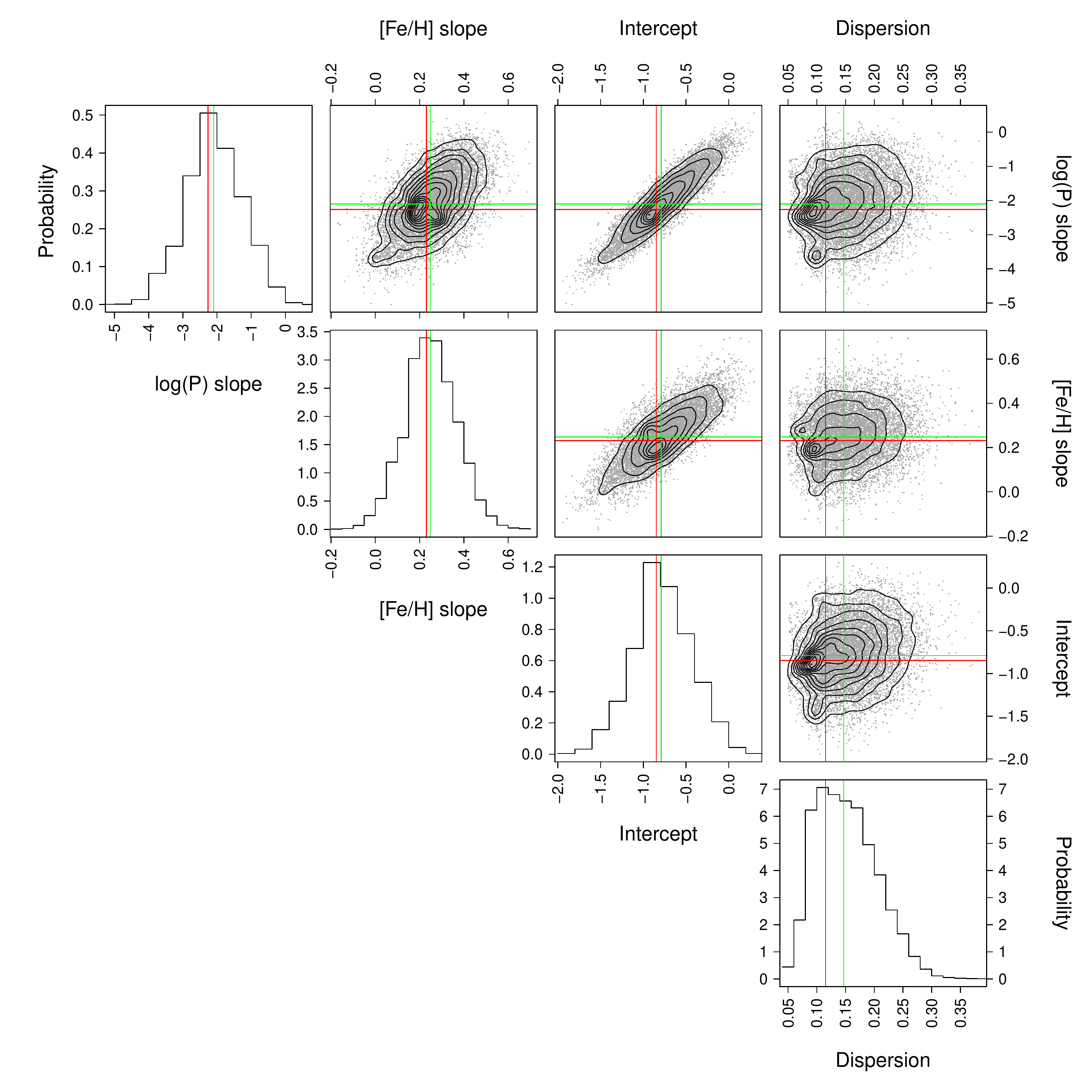}
    \caption{Marginal posterior distributions from the MCMC samples of
      the $PM_KZ$ relationship parameters for our HM { with the potential 
      TGAS parallax offset included as a model parameter}. The diagonal
      shows the uni-dimensional marginal distributions for the slope
      of the $\log(P)$ term in the linear relation, the slope of the
      metallicity term, the intercept and the intrinsic dispersion of
      the relationship. Green and red lines point respectively towards
      the median and the MAP estimate of each one-dimensional
      posterior marginal distribution.}
    \label{fig:rrl-samples}
  \end{center}     
\end{figure*}

{ Table \ref{tab:real-fitting} presents summary statistics
associated with the $PM_KZ$ relationships  obtained by our
HM trained with the RR~Lyrae sample described in Sect. \ref{sec:hiermodel}
for the cases of a potential TGAS global parallax offset inferred by the model or 
fixed to literature values. Figure \ref{fig:rrl-samples} shows the 
MCMC posterior samples (in 2D projections) of the relationship parameters 
for the model with inferred offset. The black contours in the figure represent
iso-probability lines. We see clear correlations between the three
strong parameters (two slopes and the intercept). The posterior
medians of the $\log(P)$ slope, the metallicity slope and the
intercept are, respectively, $-2.1$ mag/dex, $+0.25$ mag/dex and
$-0.79$ mag (top portion of Table \ref{tab:real-fitting}). 
The inferred period slope is consistent with the values
reported in the literature both for empirical and theoretical
studies (see Table 3 of \cite{Muraveva2015}). The metallicity slope
is systematically higher than the values reported from empirical
studies but is in good agreement with the theoretical calibrations
of \cite{Bono-2003} and \cite{Catelan-2004}.  The posterior median
of the intrinsic width is $0.15$ mag. The credible interval of the
parallax offset $\varpi_0=+0.014  \pm 0.032$ is 
in disagreement with the negative estimate $\varpi_0=-0.036 \pm 0.002$
of \cite{Arenou-et-al-2017} but is  consistent with the MAP estimate 
$\varpi_0=+0.02$ inferred by the probabilistic approach of \cite{Sesar-et-al-2017}
using $W2$-band RR Lyrae data. In any case we explain a hypothetical overtestimation
(towards positive values) of the global parallax offset inferred by our HM model  
as follows.  As we showed in Sect. \ref{sec:valid-results} the model with 3D GM prior 
mitigates the systematic overstimation of smaller parallaxes for longest periods
in a scenario of large parallax uncertainties. As a consequence of this
the lowest parallaxes inferred by the model are, on average, smaller that their
measurements (with the obvious exception of parallaxes with negative measurements),
which in turn leads to a positive offset estimate. We stress that the parallax offset
reported in this paper should no to be used as a reliable estimate of the potential
TGAS offset. On the contrary, the global parallax offset $\varpi_0=-0.057 $
reported in  \cite{paperIII} for {\textit Gaia} DR2  was inferred from  a scenario of 
sufficiently precise parallax uncertainties in which the parallax prior played a 
relatively minor role, hence, is in principle more reliable. 

 The bottom portion of Table \ref{tab:real-fitting} presents the $PLZ$
relationship parameters inferred by our HM with a global parallax offset fixed to
the value $\varpi_0=-0.036$ mas estimated by \cite{Arenou-et-al-2017}. 
We do not observe major differences in the slopes with regard to those obtained 
by the HM based on the inferred offset of $+0.014$  mas (see top portion of the table). 
Nevertheless, the difference between the intercepts is equal to 0.11  mag, which 
translates into a Large Magellanic Cloud (LMC) distance modulus $0.09$ mag longer 
than inferred by the  model with offset fixed to $\varpi_0=-0.036$  mas. 
The distance moduli have been estimated from a sample 
of 70 RRLs located close to the LMC bar, with photometry in the $K_s$-band and 
spectroscopically measured metallicities  (described and used in
 \citetalias{paperI}; \citealt{paperIII}). 
 }

\begin{table*}[htb]

  \caption {Coefficients of the $PM_KZ$ relation inferred from the
    real RRL sample described in Sect. \ref{sec:hiermodel}: slopes
    ($c$ and $k$), zero-point ($b$), intrinsic dispersion ($w$) {
      , global parallax offset ($\varpi_0$) and LMC distance modulus ($\mu_{\rm LMC}$) }.  
    The posterior distribution of each coefficient is summarized as in Table
    \ref{tab:sim-fitting-comp}. }
  \label{tab:real-fitting}

  \centering 
  \begin{tabular}{llllll}
    \hline
    \hline
    
      \multicolumn{1}{c}{$c$} & \multicolumn{1}{c}{$k$} & 
     \multicolumn{1}{c}{$b$} & \multicolumn{1}{c}{$w$}
      & \multicolumn{1}{c}{$\varpi_0$} & \multicolumn{1}{c}{$\mu_{\rm LMC}$}  \\
    \multicolumn{1}{c}{(mag/dex)}   & \multicolumn{1}{c}{(mag/dex)}  & \multicolumn{1}{c}{(mag)}    & \multicolumn{1}{c} {(mag)} & \multicolumn{1}{c} {(mas)} & \multicolumn{1}{c} {(mag)} \\

    \hline
    ${-2.10}_{-0.75}^{+0.87}$ &  ${0.25}_{-0.11}^{+0.12}$ &
    ${-0.79}_{-0.32}^{+0.36}$ & ${0.15}_{-0.05}^{+0.06}$ & ${+0.014}_{-0.032}^{+0.032}$ & ${18.73}_{-0.11}^{+0.11}$\\
    $-2.26$ & $0.23$ & $-0.85$ & $0.12$ & $+0.013$ & \multicolumn{1}{c} {$-$}\\
    
    \hline
    
    ${-2.04}_{-0.72}^{+0.77}$ &  ${0.25}_{-0.11}^{+0.11}$ &
    ${-0.68}_{-0.31}^{+0.33}$ & ${0.16}_{-0.06}^{+0.06}$ & ${-0.036}$ & ${18.64}_{-0.11}^{+0.11}$\\
    $-2.05$ & $0.25$ & $-0.71$ & $0.16$ & \multicolumn{1}{c} {$-$} & \multicolumn{1}{c} {$-$}\\
    
    \hline
  \end{tabular}	

\end{table*}

\begin{figure}
	\begin{center}
		\includegraphics[scale=0.45]{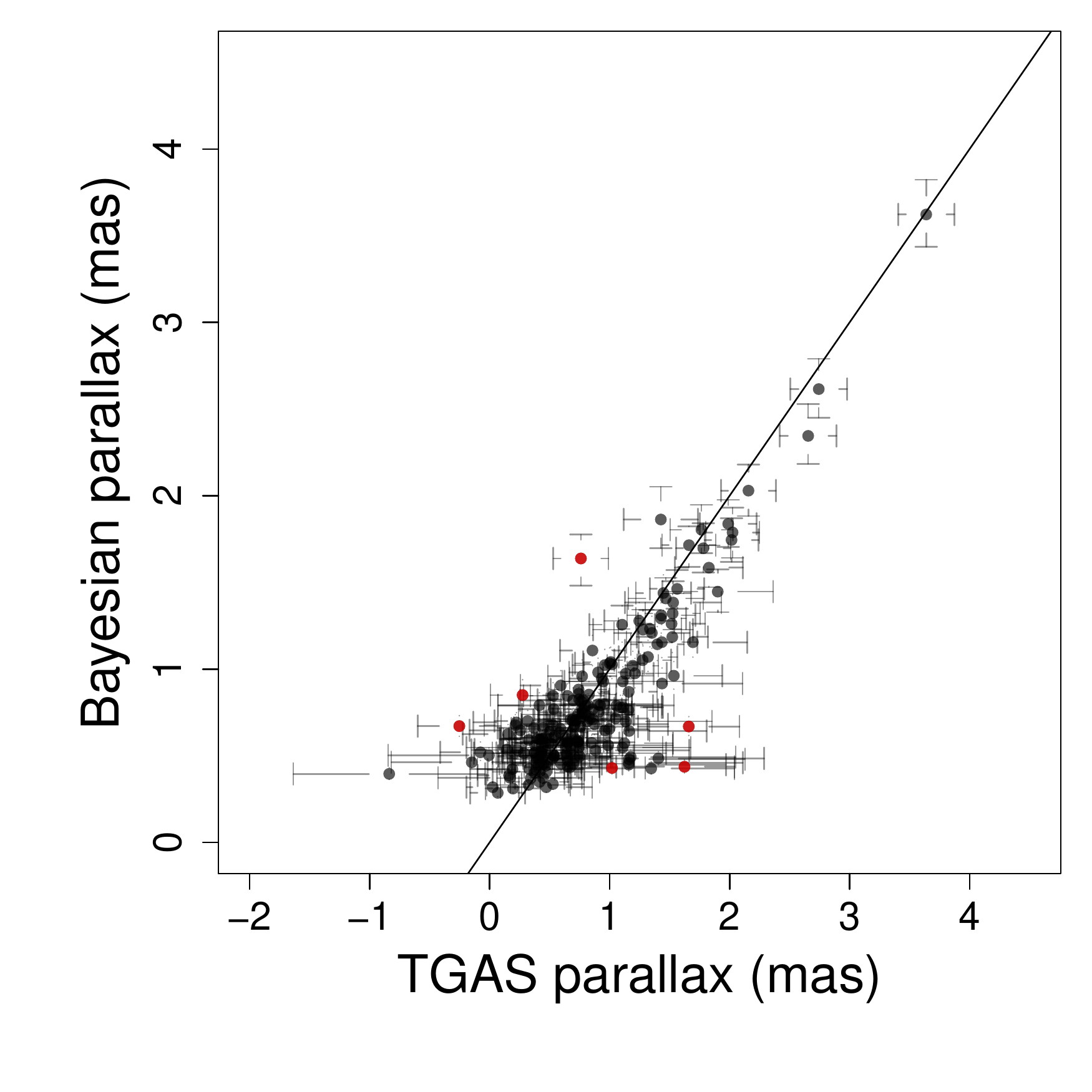}
		
		\caption{Comparison between the TGAS parallaxes and the maximum a
			posteriori estimates from the hierarchical model (HM). The error bars
			correspond to the TGAS parallax uncertainties (horizontal) and 
			credible intervals calculated as the median plus minus the difference
			between the median and the 84th and 16th percentile (vertical). 
			The red circles correspond to stars with parallax
			difference beyond 2 times the combined uncertainties.}
		\label{fig:rrl-pipi}
	\end{center}   
	
\end{figure}

{ Figure} \ref{fig:rrl-pipi} shows a comparison between the
parallaxes catalogued in TGAS and the { posterior} estimates of our
hierarchical model. { The horizontal error bars represent TGAS
  uncertainties and the vertical ones are given by 68\% credible
  intervals calculated around the median of the marginal posterior
  distributions.}  We see that our hierarchical model is capable of
reducing (``shrinking'') the uncertainties using the constraint that
the absolute magnitudes must follow a linear relationship with (the
logarithm of the) periods and metallicities with a slope { in
  agreement} with previous estimates. The { median} of the standard
deviations of the posterior samples is { 0.07} mas with a maximum
value of 0.2, which is the minimum value of the TGAS parallax
uncertainties. As shown in Fig. \ref{fig:rrl-pipi}, the maximum {
  uncertainties} of the MCMC parallax samples correspond to the same
stars with minimum TGAS uncertainties (those with maximum TGAS
parallax measurements). This means that the hierarchical model is not
capable of significantly improving the parallax uncertainties of the
stars near the Sun. We also see that there are stars with TGAS and HM
parallaxes that disagree beyond the error bars. { We plot in red
  stars that are 2-3 standard deviations away from the diagonal (as
  measured in the 2D plane of Fig. \ref{fig:rrl-pipi} by the
  Mahalanobis distance
  $\sqrt{(\mathbf{x}-\bm{\mu})^T\mathbf{\Sigma}^{-1}(\mathbf{x}-\bm{\mu})}$,
  where $\mathbf{x}$ is the vector ($\varpi_{\rm TGAS}$,$\varpi_{\rm HM}$) and
  $\bm{\mu}$ is the perpendicular projection of $\mathbf{x}$ onto the
  diagonal).}

\begin{figure*}[]
	\begin{center}
		\includegraphics[scale=0.5]{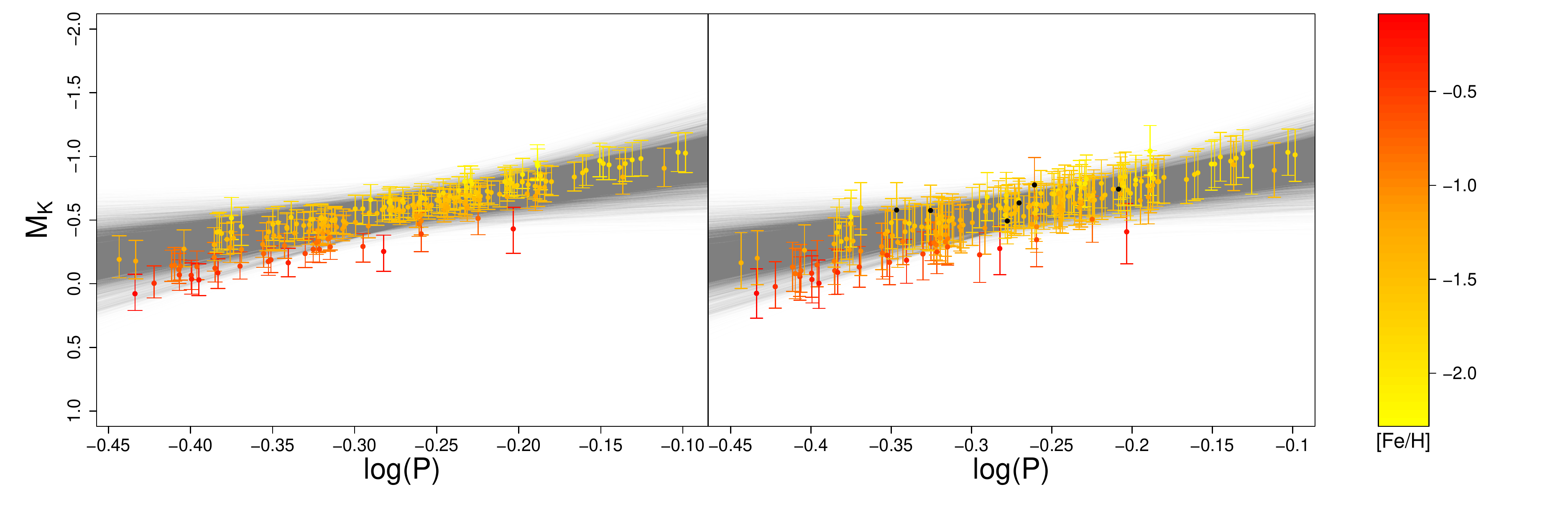}
		
		\caption{{\sl Left}. Samples of the $PLZ$ relations derived from
			the MCMC samples for ${[\rm{Fe/H}]}=-1.46$ (the median of the
			distribution of inferred metallicities) (grey lines) and
			period-$M_{\rm K}$ values inferred by the HM and computed
			according to Eq. \ref{eq:absmag1}. {\sl Right}. As in the left
			panel, but with $M_{\rm K}$ computed according to
			Eq. \ref{eq:absmag2}. The colour encodes the { inferred}
			metallicity according to the scale on the right.}
		\label{fig:PLdiagrams}
	\end{center}  
	
\end{figure*}

\begin{figure}[]
  \begin{center}
  	\hspace*{-0.5cm}
    \includegraphics[scale=0.4]{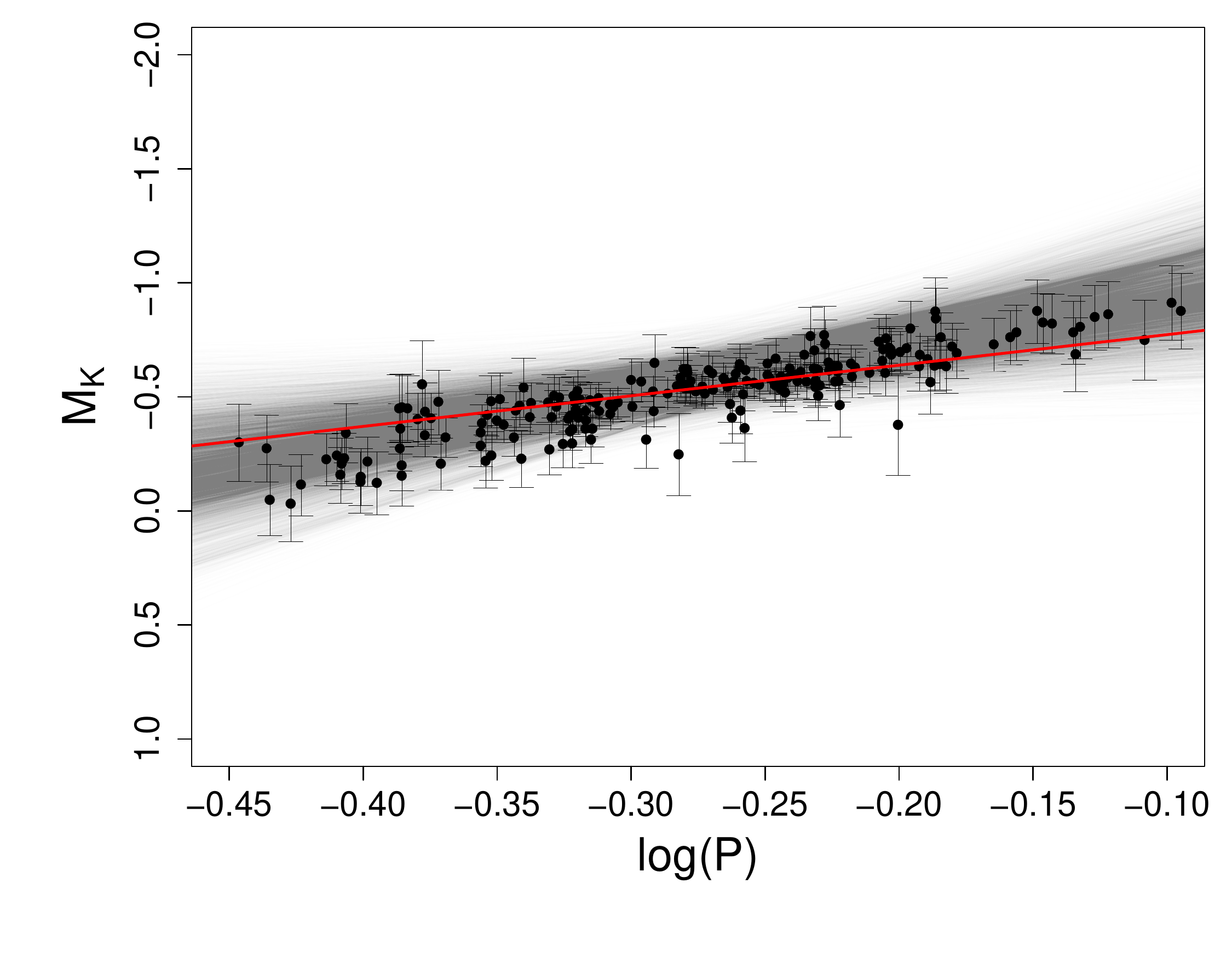}
    
    \caption{$PLZ$ relations defined by the MCMC samples for the value
      of ${[\rm{Fe/H}]}=-1.46$ (dark grey lines) and the measured
      periods, and absolute magnitudes predicted by using the ABL method
       described in the text. The red line represents the projection of
      the $PLZ$ relation used for predictions for a value of the metallicity
      equal to its median in the sample.} 
    
    \label{fig:meas}
  \end{center}
\end{figure}

{ Figure} \ref{fig:PLdiagrams} shows the PL relations derived from
the HM. Each grey line corresponds to one sample in the Markov
chain. All $PM_KZ$ relations have been particularized to a value of
the metallicity ${[\rm{Fe/H}]}=-1.46$ dex, which is the median of the
distribution of { inferred} values. On the left hand panel we show
the values of the absolute magnitude in the $K$ band derived from the
MCMC samples as

\begin{equation}
  M_{\rm K}^{i,n} = c^{n}\cdot\log(P^{i,n})+k^{n}\cdot
  {[\rm{Fe/H}]}^{i,n}+b^{n}\,,
  \label{eq:absmag1}
\end{equation}
   
\noindent where the superindex $i$ tags stars (from 1 to $N$) and the
superindex $n$ tags the sample in the MCMC set of samples. In the
right hand panel we show the same diagram, but computing the absolute
magnitude from the measured apparent magnitude and the MCMC parallax:
   \begin{equation}
   M_{\rm K}^{i,n} = \hat{m}_0^i+5\cdot\log(\varpi^{i,n})-10\,,
   \label{eq:absmag2}
   \end{equation}

\noindent where $\hat{m}_0^i$ represents the measured value of the
apparent magnitude corrected for the measured { absorption} (that
is, we only use the parallaxes from the model, but the {
  absorptions} and apparent magnitudes used in Eq. \ref{eq:absmag2}
are the { measured values in the sample described in
  Sect. \ref{sec:hiermodel}} .  The black circles correspond to the
discrepant sources marked by red circles in
Fig. \ref{fig:rrl-pipi}. The outlier at $\log(P)\approx -0.26$
corresponds to V363 Cas. This star was classified as a double-mode
pulsator by \cite{Hajdu-et-al-2009}. A detailed analysis of its
nature is beyond the scope of this paper, but we note that it would be
consistent with its discrepant position in the diagrams. 
{ The two panels of  Fig. \ref{fig:PLdiagrams} can be compared 
to the period-absolute magnitude diagram of Fig. \ref{fig:meas} in which 
the absolute magnitudes were predicted from 
a $PLZ$ relation whose coefficients  were estimated fitting by
 weighted non-linear least squares the following model  
\begin{equation}
\hat{\varpi}10^{0.2\hat{m}_{0}-2}=10^{0.2\left(c\log\hat{P}+
	k\left[\widehat{\mathrm{Fe/H}}\right]+b\right)}\,,
\label{eq:ABL}
\end{equation}
where the dependent variable $\alpha=\hat{\varpi}10^{0.2\hat{m}_{0}-2}$ in
the left side of Eq. \ref{eq:ABL} is the Astrometry-Based Luminosity
(ABL) defined by \cite{Arenou-Luri-1999} with $\hat{\varpi}$ denoting the TGAS
parallax. The red line in Fig. \ref{fig:meas} represents the projection of
the $PLZ$ relation derived by this method for a value of the metallicity
equal to its median in the sample. The $\log(P)$ slope, metallicity slope 
and intercept estimates were respectively $c= -1.34 \pm 0.95$ mag/dex, 
$k= 0.20 \pm 0.13$ mag/dex and $b= -0.62 \pm 0.40$ mag.  

}

\begin{figure}
  \begin{center}
    \includegraphics[scale=0.55]{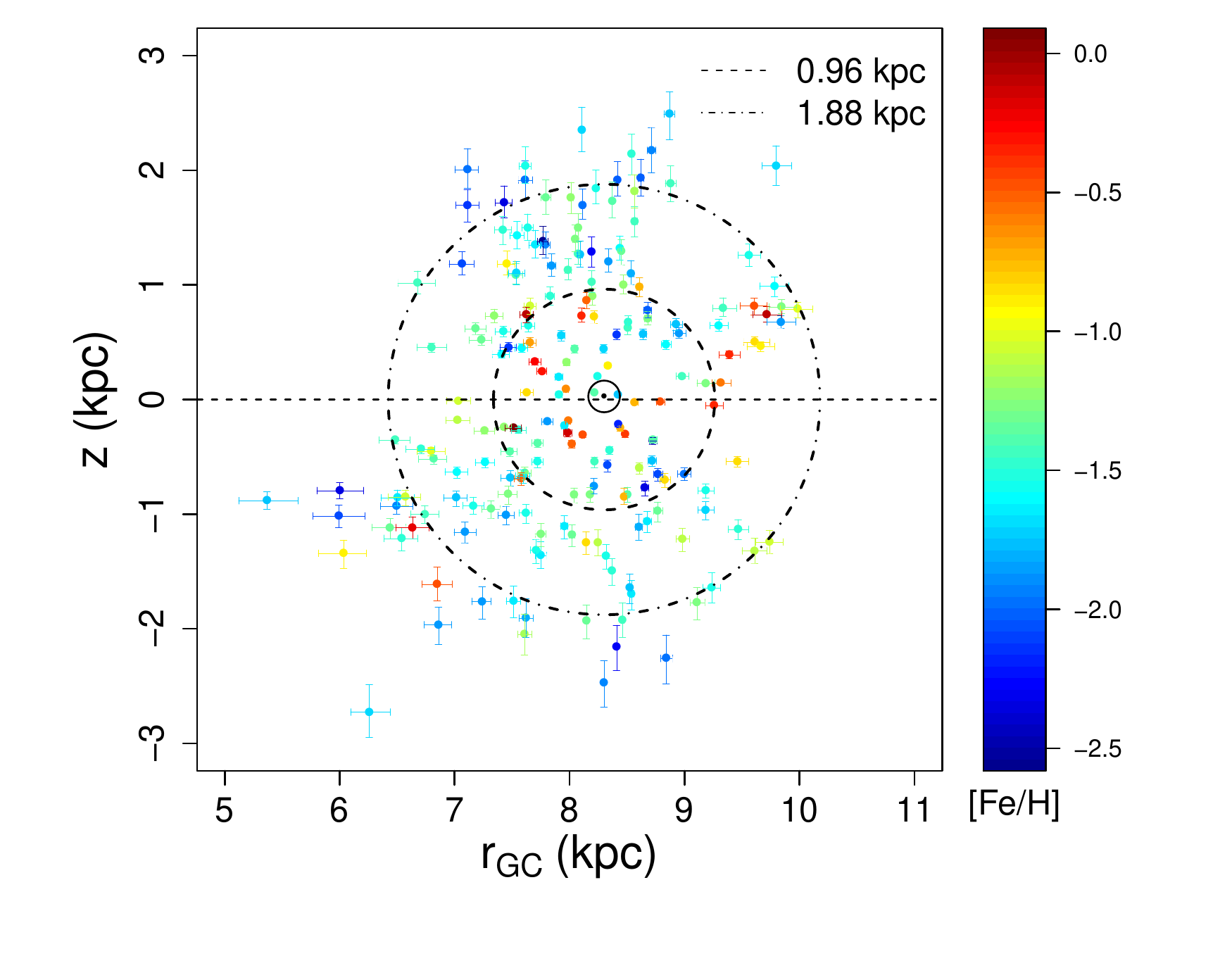}
		
    \caption{Spatial distribution and metallicity of the RR~Lyrae sample
      used in this paper represented in a Galactocentric reference
      frame.  $r_{\rm GC}$ and $z$ denote respectively the radial
      distance to the Galactic centre (GC) and the vertical distance
      with regard to the Galactic midplane $x-y$. The Cartesian
      Galactocentric coordinates and the radial distances to the GC 
      have been summarized by the median plus
      minus the difference in absolute value between the median and the
      84th and 16th percentile of the posterior samples
      of Eqs. \ref{eq:GC.coord} and \ref{eq:r.GC}.
      }
	\label{fig:rrl-galact}
	\end{center}   
	
\end{figure}

{ One of the advantages of addressing the problem of calibrating a
  $PL(Z)$ relationship by means of a Bayesian HM is the ability to
  infer the posterior distribution of any parameter of interest.  In
  particular we aimed to derive individual heliocentric distances to
  the RRL stars in our sample and locate their positions in the
  Galaxy.  For that, the measured coordinates $(\alpha,\delta)$ of our
  RRL stars were first transformed to Galactic coordinates
  $(l,b)$. Then, for each RR~Lyrae we derived samples of the posterior
  distribution of its rectangular coordinates $(x_i,y_i,z_i)$ from 
  samples of its posterior parallax ($\varpi_i$) in the Cartesian
  Galactocentric coordinate system of \cite{Juric-2008} by
  
  \begin{equation}
  \begin{split}
  x_{i}^{n}&=R_\odot-d_{i}^{n}\cos\left(l_{i}\right)\cos\left(b_{i}\right)\\
  y_{i}^{n}&=-d_{i}^{n}\sin\left(l_{i}\right)\cos\left(b_{i}\right)\\
  z_{i}^{n}&=d_{i}^{n}\sin\left(b_{i}\right) \,,\\
  \end{split}
  \label{eq:GC.coord}
  \end{equation}
  
\noindent where $d_{i}^{n}=1/\varpi_{i}^{n}$ denotes the posterior distance 
calculated as the reciprocal of the posterior parallax and $R_\odot=8.3$ kpc is the adopted
distance to the Galactic centre \citep{Gillessen-2009}. The samples of the
posterior distribution of the radial distance of each star to the Galactic centre
were calculated as

\begin{equation}
	{r_{\mathrm{GC}}}_{i}^{n}=\sqrt{\left(x_{i}^{n}\right)^{2}+
		\left(y_{i}^{n}\right)^{2}}\,.
	\label{eq:r.GC}
\end{equation}

 Figure \ref{fig:rrl-galact} represents the spatial distribution and 
 metal abundance of our RR~Lyrae sample in the plane $z-r_{\mathrm{GC}}$  
 associated with the Galactocentric reference frame.  Each Sun-centred circumference 
 in the figure corresponds to Galactocentric coordinates calculated from 
 the Galactic longitudes  $l=0^\circ$ and $l=180^\circ$ (at any Galactic latitude) 
 and the distances corresponding to the median of the mean log-parallax 
 posterior distribution associated with each Gaussian mixture component of our HM.

 }

 \begin{table*}[htb]
 	\caption{ Summary statistics corresponding to the sensitivity
 		analysis performed to the HM with joint 3D GM prior presented
 		in this paper: slopes ($c$ and $k$), zero-point ($b$), intrinsic dispersion ($w$)  
 		and global parallax offset ($\varpi_0$).
 		The posterior distribution of each coefficient is summarized as in Table
 		\ref{tab:sim-fitting-comp}. }
 	\label{tab:SA-stats}
 	\centering
 	\begin{tabular}{llllll}

 		Prior  & \multicolumn{1}{c}{$c$} & \multicolumn{1}{c}{$k$} & 
 		\multicolumn{1}{c}{$b$} & \multicolumn{1}{c}{$w$}  & \multicolumn{1}{c}{$\varpi_0$}   \\
 		& \multicolumn{1}{c}{(mag/dex)}   & \multicolumn{1}{c}{(mag/dex)}  & \multicolumn{1}{c}{(mag)}    
 		& \multicolumn{1}{c} {(mag)} & \multicolumn{1}{c}{(mas)}   \\
 		
 		\hline
 		\hline 
 		
 		$\pi\left(c \right)=\pi\left(k \right)=\mathsf{Cauchy}\left(0,\sigma=2.5\right)$ & $-2.23 \substack{+0.78\\-0.77}$ 
 		& $+0.24\substack{+0.12\\-0.11}$   & $-0.83\substack{+0.33\\-0.34}$
 		& $0.16\substack{+0.06\\-0.05}$  & $+0.015\substack{+0.035\\-0.035}$ \\
 		
 		&$-2.22$ & $+0.24$ & $-0.83$ &  $0.15$ & $+0.015$ \\
 		
 		\hline 
 		
 		$\pi\left(w\right)=\mathsf{Exp}\left(\lambda=0.1\right)$ & $-2.05\substack{+0.81\\-0.81}$ 
 		& $+0.26\substack{+0.11\\-0.11}$   & $-0.75\substack{+0.35\\-0.35}$
 		& $0.16\substack{+0.05\\-0.06}$ & $+0.012\substack{+0.035\\-0.033}$  \\
 		
 		&$-2.06$ & $+0.26$ & $-0.74$ &  $0.15$ & $+0.008$ \\
 		
 		\hdashline	
 		
 		$\pi\left(w\right)=\mathsf{Exp}\left(\lambda=1\right)$  & $-2.10\substack{+0.87\\-0.75}$  
 		& $+0.25\substack{+0.12\\-0.11}$ & $-0.79\substack{+0.36\\-0.32}$ 
 		& $0.15\substack{+0.06\\-0.05}$  & $+0.014\substack{+0.032\\-0.032}$ \\
 		
 		&$-2.26$ & $+0.23$ & $-0.85$ &  $0.12$ & $+0.013$ \\
 		
 		\hdashline
 		
 		$\pi\left(w\right)=\mathsf{Exp}\left(\lambda=10\right)$ & $-2.10\substack{+0.76\\-0.78}$ 
 		& $+0.25\substack{+0.11\\-0.11}$   & $-0.79\substack{+0.32\\-0.33}$ 
 		& $0.12\substack{+0.05\\-0.04}$ & $+0.015\substack{+0.032\\-0.032}$  \\
 		
 		&$-2.13$ & $+0.23$ & $-0.69$ &  $0.12$ & $+0.016$ \\
 		
 		\hline 	
 		
 		$\pi\left(\boldsymbol{\mu}^{k}\right)$ with $\sigma_{\mu_{\varpi}}^{k}=0.1$ &  $-1.92\substack{+0.81\\-0.69}$   
 		& $+0.21\substack{+0.11\\-0.10}$   & $-0.63\substack{+0.35\\-0.30}$   
 		& $0.16\substack{+0.06\\-0.05}$ & $-0.006\substack{+0.039\\-0.040}$ \\
 		
 		&$-2.08$ & $+0.20$ & $-0.78$ &  $0.15$ & $-0.008$ \\	
 		
 		$\pi\left(\boldsymbol{\mu}^{k}\right)$ with $\sigma_{\mu_{\varpi}}^{k}=0.5$ &  $-1.60\substack{+0.81\\-0.69}$   
 		& $+0.17\substack{+0.11\\-0.11}$   & $-0.64\substack{+0.32\\-0.33}$  
 		& $0.17\substack{+0.06\\-0.05}$ & $-0.023\substack{+0.043\\-0.046}$ \\	
 		
 		&$-1.58$ & $+0.16$ & $-0.64$ &  $0.17$ & $-0.021$ \\
 		
 		\hline
 		
 		\hline
 	\end{tabular}
 \end{table*}

\section{Sensitivity analysis}
\label{sec:sensi}

{ In this section we analyse the sensitivity of the inference
  results obtained by our HM with a joint 3D GM prior for metallicity,
  period and parallax to variations of some critical parameters
  assigned to the different prior distributions. In the following, we
  consider the hyperparameters of Table \ref{tab:priors} (for which
  the results presented in Sect. \ref{sec:post} were obtained) as
  reference values and compare these results with those obtained varying
  the values of the hyperparameters. Table \ref{tab:SA-stats} compares
  the posterior medians and credible intervals of the $PLZ$
  relationship parameters for the different values of the prior
  hyperparameters. Its third row lists the reference results
  introduced in Sect. \ref{sec:post}.
  
In Sect. \ref{sec:hiermodel}, the prior distributions of the $PLZ$
relationship coefficients were chosen, for obvious reasons, to be as
non-informative as possible. For the sample of RRL stars analysed in
this paper we expect significant variations of the posterior
distributions for other choices of their prior hyperparameters because
in this range of uncertainties, the prior plays a major role.  In
particular the assignment of the slightly more informative popular
Cauchy prior with scale parameter $\sigma=2.5$ of \cite{Gelman-2008}
to the slopes $c$ and $k$ gave rise to a larger absolute value of the
$\log(P)$ slope posterior median and a slightly narrower credible
interval (top portion of Table \ref{tab:SA-stats}).  The closeness of
the posterior median and MAP estimates indicates that the MCMC
algorithm explored successfully the complex parameter space of the
problem in this case. 

We have also tried different values for the inverse scale
hyperparameter $\lambda_w$ of the intrinsic width prior distribution
from 0.1 to 10 kpc. The results (middle portion of the table) indicate
a slight decrease of the intrinsic dispersion as $\lambda_w$
increases.

The most critical HM hyperparameters are those that correspond to the
GM prior of Eq. \ref{eq:P-Z-Pi-dist} which models the true
distribution of metallicities, periods and parallaxes. We have
assigned informative hyperparameters to the standard deviations
$\left(\sigma_{\mu_{Z}}^{k},\sigma_{\mu_{P}}^{k},\sigma_{\mu_{\varpi}}^{k}\right)$
of the 3D Gaussian prior chosen for the mean vector
$\boldsymbol{\mu}^{k}$ of each GM component. In particular, for the
logarithm of parallax we chose $\sigma_{\mu_{\varpi}}^{k}=0.05$. We
have also tested the values $0.1$ and $0.5$ with the results listed in
the bottom portion of Table \ref{tab:SA-stats} where we observe that
the $\log(P)$ slope is severely underestimated for
$\sigma_{\mu_{\varpi}}^{k}=0.5$. For this latter case the inferred
posterior medians of ${\mu_{\varpi}}$ were equal to $-0.50$ and
$-0.04$ (equivalent to $0.60$ and $0.95$ mas or $1.65$ and $1$ kpc).
These are the only cases in which the inferred parallax offset 
turns out to be negative and reflect the fact that the 3D GM prior 
is not constricting adequately the range of smaller parallaxes.  
}

\section{Summary and conclusions}
\label{sec:summary}
   
In this paper we have applied the hierarchical Bayesian methodology to
infer { estimates} for the parameters of the $PLZ$ relationship in
the $K$-band for fundamental and first overtone RR Lyrae stars. We
have extended the analysis performed in \citetalias{paperI} by testing
new prior distributions and analysing correlations in the data, their
influence on the inference and the consequences of the prior choice.
   
In Sect. \ref{sec:validation} we have demonstrated through the use of
semi-synthetic data that the RR Lyrae sample used in
\citetalias{paperI} presents strong correlations that result in
different spatial distributions for the different metallicites and
periods. As a result, the larger parallax uncertainties are not spread
uniformly in period but concentrated in the region of long periods,
thus making the inference results strongly dependent on the
prior. { This is the main result of this work. We prove that in the
  context of significant parallax uncertainties (in the TGAS samples,
  this amounts to a median fractional uncertainty
  $\sigma_{\varpi}/\varpi$ of 0.43), simple independent priors will
  result in systematically biased estimates of the PLZ slopes and
  intercept. For small parallax uncertainties (typically one order of
  magnitude smaller than the TGAS uncertainties) the effect of such
  correlations on the parameters of the $PLZ$ relation inferred by our
  HM under a wide variety of priors is small. In such simulated
  scenario, our HM is able to successfully recover the $PLZ$ relation
  of \cite{Catelan-2004} given by Eq. \ref{eq:PKZ-Catelan-2004}
  independently of the prior choice. On the contrary, for the TGAS
  parallax uncertainties used in \citetalias{paperI} we propose a
  Mixture of two Gaussian 3D components with correlations between
  periods and parallaxes. We prove that this prior is much less
  affected by the correlations in the data set and that it recovers
  the right parameters for semi-synthetic data with metallicity
  uncertainties that are half of those available in the
  literature. This (the inadequacy of 1D priors for the inference of
  PL(Z) relations and the necessity of modelling the correlations in
  the data set) is the second main result of our study.}
     
{ In Sect. \ref{sec:post} we have applied our HM to the sample of
  200 fundamental and first overtone RR Lyrae stars and Gaia DR1
  parallaxes used in \citetalias{paperI}. The value of the $PM_KZ$
  coefficients thus derived ($c=-2.10_{-0.75}^{+0.87}$,
  $k={0.25}_{-0.11}^{+0.12}$, $b={-0.79}_{-0.32}^{+0.36}$, and
  $w={0.15}_{-0.05}^{+0.06}$) can be compared with the estimate
  derived from the much more precise measurements of the {\textit
    Gaia} second Data Release \citep[$c=-2.58_{-0.20}^{+0.20}$,
    $k={0.17}_{-0.03}^{+0.03}$, $b={-0.84}_{-0.09}^{+0.09}$, and
    $w={0.16}_{-0.01}^{+0.01}$;][]{paperIII}. We see that the 68\%
  credible intervals have large overlap regions making the two
  estimates fully consistent. We note that the results presented in
  \cite{paperIII} already incorporate the findings of the study
  presented here with the only exception that the 3D prior for
  parallaxes, periods and metallicities is not a mixture of Gaussians
  but a single Gaussian distribution because in the {\textit Gaia} DR2
  typical parallax uncertainty regime the data are sufficiently
  precise to disentangle the two metallicity populations without
  enforcing this separation in the prior.}

\bibliographystyle{aa}
\bibliography{references_m}

\end{document}